\documentclass[12pt]{article}
\usepackage{geometry}
\usepackage[dvipsnames]{xcolor}
\usepackage{amsmath}
\usepackage{amssymb}
\usepackage{dsfont}
\usepackage[english]{babel}
\usepackage[utf8]{inputenc}
\usepackage{times}
\usepackage{graphics}
\usepackage{graphicx}
\usepackage[T1]{fontenc}
\usepackage[official,right]{eurosym}
\usepackage{url}
\usepackage{eso-pic}
\usepackage{fancyhdr}
\usepackage{graphicx}
\usepackage{subcaption}
\usepackage{adjustbox}
\usepackage{booktabs}

\usepackage[utf8]{inputenc}
\usepackage{graphicx}
\graphicspath{{Images/}}

\usepackage{natbib}
\usepackage{tabularx}
\usepackage{lscape}
\usepackage{mathtools}
\usepackage{changepage}
\usepackage{multirow}
\usepackage{amsmath}
\usepackage{tensor}
\usepackage{pgf}
\usepackage{pgfplots}
\usepackage{amssymb}
\usepackage{mathtools,zref-savepos}
\usepackage{authblk}
\usepackage{version}
\usepackage{float}
\usepackage{changepage}
\usepackage{rotating}
\usepackage{tikz}
\usepackage{amsmath}
\usepackage{multirow}
\usepackage{graphicx}
\usepackage{comment}
\usepackage{setspace}
\usepackage[ruled,lined]{algorithm2e}

\pgfplotsset{compat=1.18}
\newtheorem{prop}{Proposition}

\newtheorem{ax}{Axiom}
\newtheorem{ex}{Example}
\newtheorem{lemma}{Lemma}

\usetikzlibrary{plotmarks}
\usetikzlibrary{patterns}
\usetikzlibrary{shapes}
\usepackage{minitoc}
\usepackage[normalem]{ulem}

\definecolor{green}{HTML}{34A853}
\definecolor{blue}{HTML}{4285F4}
\definecolor{yellow}{HTML}{FBBC05}
\definecolor{red}{HTML}{EA4335}

\DeclareMathAlphabet{\pazocal}{OMS}{zplm}{m}{n}

\newcommand{\A}{\mathbf{A}}
\newcommand{\B}{\mathbf{B}}
\newcommand{\C}{\mathbf{C}}
\newcommand{\D}{\mathbf{D}}

\newcommand{\dm}{\mathcal{D}}

\newcommand{\ssk}{\smallskip}
\newcommand{\msk}{\medskip}

\newcommand{\wrt}{w.r.t.\ }

\newcommand{\ie}{i.e.,\xspace}

\newcommand{\wl}{w.l.o.g.\xspace}

\newcommand{\epr}{\hfill $\blacksquare$\mbox{} }
\newcommand{\bpr}{\noindent {\bf Proof.} \hspace{1 em}}

\newtheorem{defn}{Definition}[section]
\newtheorem{cor}{Corollary}

\newcommand{\Rhnneg}{\ensuremath{\mathbb{R}^{h^*}_{\geq 0}}}
\newcommand{\Rhpos}{\ensuremath{\mathbb{R}^{h^*}_{\geq0}}}
\newcommand{\nsdom}[1][\A]{\ensuremath{P(#1)}}

\newcommand{\ovx}[1][x]{\overline{#1}}

\title{On compromising freedom of choice and
subjective value.}
\author{Nicolas Fayard$^{\dagger}$, Marc Pirlot$^{\ddagger}$, Alexis Tsoukiàs$^{\dagger}$\\$\dagger$ CNRS-LAMSADE, PSL, Université Paris Dauphine\\
$\ddagger$MATHRO, Faculté Polytechnique, Université de Mons}
\date{November 2025}

\begin{document}

\thispagestyle{empty}

\enlargethispage*{8cm}
 \vspace*{-38mm}

\AddToShipoutPictureBG*{\includegraphics[width=\paperwidth,height=\paperheight]{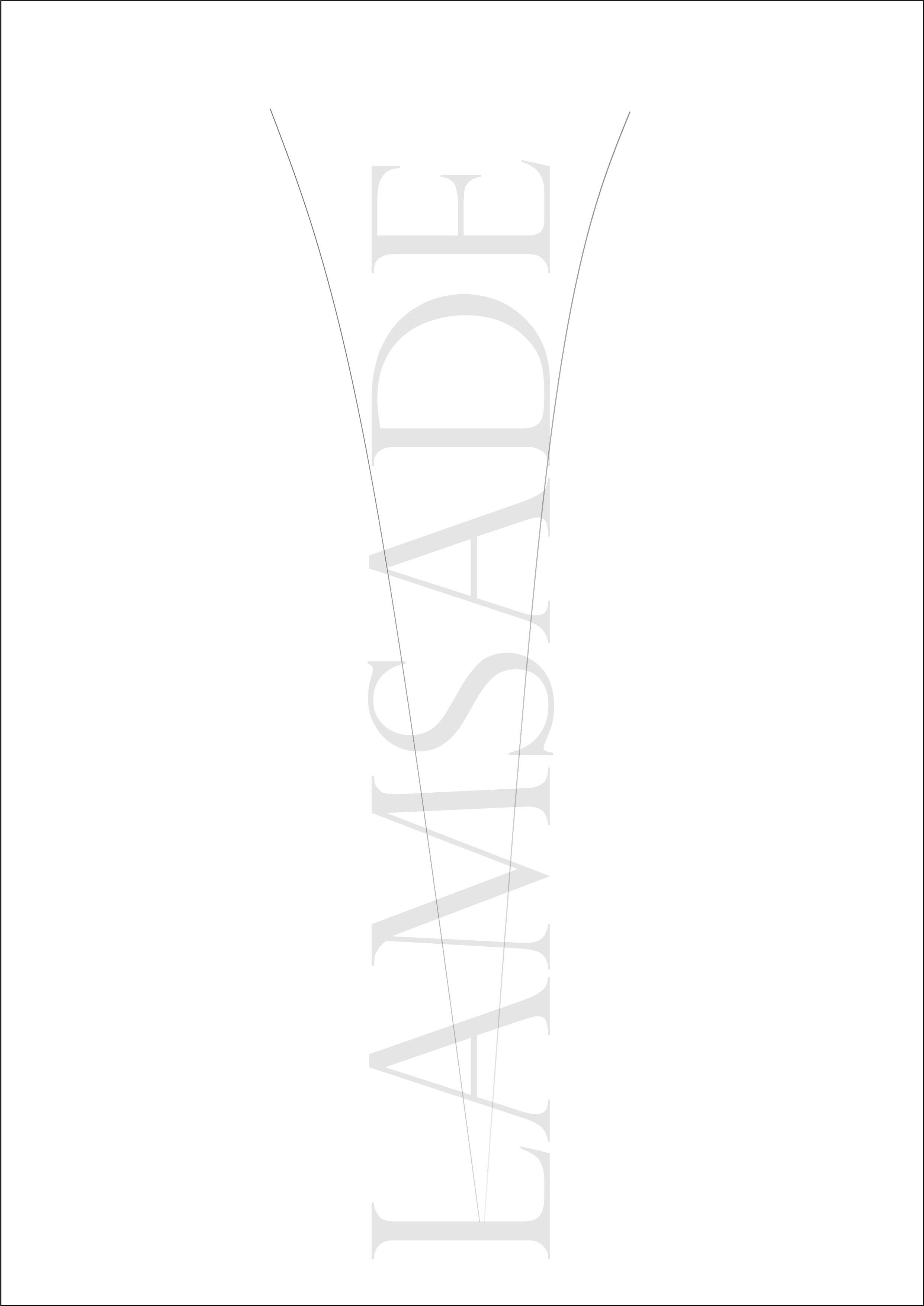}}

\begin{minipage}{24cm}
 \hspace*{-28mm}
\begin{picture}(500,700)\thicklines
 \put(60,670){\makebox(0,0){\scalebox{0.7}{\includegraphics{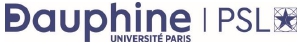}}}}
 \put(60,70){\makebox(0,0){\scalebox{0.3}{\includegraphics{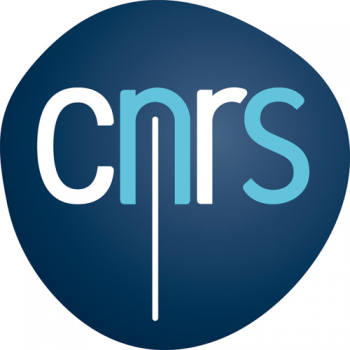}}}}
 \put(320,350){\makebox(0,0){\Huge{CAHIER DU \textcolor{BurntOrange}{LAMSADE}}}}
 \put(140,10){\textcolor{BurntOrange}{\line(0,1){680}}}
 \put(190,330){\line(1,0){263}}
 \put(320,310){\makebox(0,0){\Huge{\emph{412}}}}
 \put(320,290){\makebox(0,0){November 2025}}
 \put(320,210){\makebox(0,0){\Large{On compromising freedom of choice and
subjective value.}}}
 \put(320,190){\makebox(0,0){\Large{subtitle}}}
 \put(320,100){\makebox(0,0){\Large{Nicolas Fayard, Marc Pirlot, Alexis Tsoukiàs}}}
 
 \put(320,670){\makebox(0,0){\Large{\emph{Laboratoire d'Analyse et Mod\'elisation}}}}
 \put(320,650){\makebox(0,0){\Large{\emph{de Syst\`emes pour l'Aide \`a la D\'ecision}}}}
 \put(320,630){\makebox(0,0){\Large{\emph{UMR 7243}}}}
\end{picture}
\end{minipage}

\newpage

\addtocounter{page}{-1}

\maketitle

\abstract{This paper proposes a new framework for evaluating capability sets by incorporating individual preferences over the diversity of accessible options. Building on the Capability Approach, we introduce a compromise method that balances between the notions of negative and positive freedom, effectively capturing the intrinsic and instrumental values of diverse choices within capability sets.}

\newpage

\section{Introduction}
The Capability Approach, introduced by Amartya \cite{sen1979equality,sen1985commodities,sen1993capability,sen1999development,sen2009idea}, provides a framework for evaluating individual well-being and social justice by shifting the focus from material resources to the real freedoms people have to achieve the lives they value. Central to this approach is the concept of capability set, which represents the set of achievable functionings (seen as beings and doings) that individuals can choose from, given their resources, personal characteristics, and the surrounding environment. This perspective ties freedom directly to the availability and quality of opportunities, emphasizing that freedom of choice is not only about the range of options but also about their ability to make individuals lead meaningful lives.

\ssk The link between opportunity set theory and the Capability Approach lies in their shared focus on freedom as a central metric for well-being. Opportunity set theory provides a formal framework for comparing sets of choices available to individuals, capturing the idea that freedom can be understood as a range of opportunities. This aligns directly with the Capability Approach, which interprets these sets as reflecting the real opportunities individuals have to achieve valued outcomes. By analyzing the structure, size, and relevance of opportunity sets, one can formalize and measure the extent of freedom in a way that resonates with the normative goals of the Capability Approach \citep{pattanaik1990ranking,xu2003functionings,kreps1979preference,carter1995measure}.

In the Capability Approach, the emphasis is not on the range of nominal alternatives, but on the real opportunities to achieve valuable states of being. Each option is conceived as a vector of functionings across several welfare dimensions, and a capability set gathers the attainable vectors. 
From this perspective, adding a dominated element does not enlarge what a person can truly achieve, which makes it natural to focus on the set of undominated elements (the Pareto frontier). 
This stance is normatively stronger than in some opportunity set models \citep{pattanaik1990ranking,gaertner2008new}, where dominated options may still be considered relevant. For this reason, we will use the term \emph{capability sets} rather than \emph{opportunity sets}.

\ssk 
Within the broader discussion of freedom, scholars have highlighted two interconnected aspects: interpretations of freedom and methods to measure it. The interpretations, such as negative freedom, positive freedom, and preference for flexibility (see below for explanations and references), offer conceptual foundations for understanding the value of capability sets. These perspectives emphasize distinct but overlapping ways in which freedom manifests in capability sets. Meanwhile, the development of measurement approaches aims to operationalize these interpretations by providing quantitative tools to evaluate the extent and quality of freedom. Together, these two aspects form the backbone of opportunity set theory and its practical applications.

\ssk Freedom interpretations have been categorized into three main perspectives. Negative freedom, rooted in the absence of external constraints, views freedom as the ability to act without interference, focusing on the size of opportunity sets \citep{berlin1969four,pattanaik2002minimal,oppenheim1981dimensions}. Negative freedom is closely linked to the intrinsic value of freedom, as it emphasizes the existence of options and the autonomy they represent, regardless of whether these options are utilized. Positive freedom emphasizes the availability and conversion of meaningful opportunities, highlighting the quality of options \citep{sen1985commodities,carter2004choice}. Positive freedom is inherently tied to the instrumental value of freedom, as it focuses on how opportunities enable individuals to achieve valuable outcomes and lead fulfilling lives. Finally, preference for flexibility, introduced by  \cite{kreps1979preference}, stresses the adaptability provided by larger sets in the face of future preference uncertainty \citep{carter1995measure,vanhees2000legal}.

\ssk The methods for measuring those sets of options have evolved to capture these varying interpretations. Quantity based methods \citep{pattanaik1990ranking,pattanaik2000ranking}, with the simplest approach, counts the number of options in a set, aligning closely with negative freedom. Weighted cardinality refines this by assigning value to each option based on its significance, connecting with positive freedom \citep{xu2003functionings,carter2004choice,pattanaik1998preference,xu2003functionings,xu2003ranking}. Utility-based measures evaluate sets by considering the maximum utility achievable, often linked to both positive freedom and preference for flexibility \citep{kreps1979preference,pattanaik2015freedom}. Finally, measures of diversity focus on the range and variation of choices, addressing elements of all three interpretations by highlighting the richness and adaptability of the set \citep{carter1995measure,xu2003functionings}.

\ssk In this paper, we position our work in between negative and positive freedom. Our proposed measure is situated between utility-based approaches and diversity-based approaches. To the authors' knowledge, the only works that have attempted to measure freedom as conjointly reflecting the diversity and the valuation of options are \cite{gaertner2012evaluating,gaertner2006capability,gaertner2008new,gaertner2011reference}. However, a key limitation of their methods is that they do not satisfy the Strong Monotonicity Property, meaning that adding an alternative that appears relevant to an individual does not necessarily increase their freedom.

\ssk Section 2 introduces the mathematical setting and the  notation. Section 3 delves into the necessity of developing a function to assess capabilities, focusing on the role of freedom of choice and exploring the Gaertner and Xu methods and the volume based method proposed by \cite{xu2004ranking}. Section 4 defines desirable properties of methods for assessing capability sets. Section 5 presents our approach, the compromise approach, which respects those properties. Our approach will be compared to two methods that represent opposing perspectives on the value of diversity in capability sets. The \emph{instrumental extreme} evaluates an opportunity set only by its best-valued option, reflecting a purely outcome-oriented view of freedom. In contrast, the \emph{intrinsic extreme} values freedom through the availability of all equally desirable alternatives, emphasizing the richness of choice. These two serve as normative extremes between which our proposed compromise approach can be flexibly parameterized. Proofs of the main results are provided in the appendix.


\section{A formal framework for Capability sets}\label{Notation}

An individual's being is represented by a vector $\vec{a} = (a_1, \ldots, a_h, \ldots, a_{h^*}) \in \Rhpos = \{\vec{x} \in \mathbb{R}^{h^*} | x_h \geq 0, \forall h= 1, \ldots h^*\}$, where the coordinates $a_h$ assess the performance levels attained on a set of $h^*$ welfare dimensions (or functionings), such as life expectancy, level of nourishment, mobility, etc. It is assumed that the evaluations on the $h^*$ dimensions of the space $\Rhpos$ provide a complete description of the welfare aspects that are relevant for the individual. We also assume, \wl, that, for all welfare dimensions, the larger the evaluation the better.

\ssk We recall the definition of the weak (resp. strict) dominance relation on $\Rhpos$. For $\vec{a},\vec{b} \in \Rhpos$, we have that $\vec{a}$ weakly dominates $\vec{b}$ (denoted $\vec{a} \geq \vec{b}$) if $a_h \geq  b_h$, for all $h = 1, \ldots, h^*$. We say that $\vec{a}$ strictly dominates $\vec{b}$ (denoted $\vec{a} > \vec{b}$) if $\vec{a}$ weakly dominates $\vec{b}$ and $a_h > b_h$, for some $h \in \{1, \ldots, h^*\}$\footnote{We abuse notation by using the same symbol $\geq$ for denoting the usual order relation on the real numbers set $\mathbb{R}$ and the weak dominance relation on $\Rhpos$. A similar remark holds for the symbol $>$.}. We also use the notation $\leq$ and $<$ in an obvious way. Under the above made assumption, \ie that larger evaluations are better than smaller for all welfare dimensions, a being that dominates another is necessarily preferred to the latter.

In practice, a capability set can take different forms depending on the problem under study.  
It may be a finite collection of attainable beings, for example when an empirical dataset enumerates distinct combinations of functionings that a person can realistically achieve \citep{Fayard2024}.  
In other contexts, it may be an infinite set defined implicitly by constraints,  such as the individual’s limited resources, together with a model describing how these resources can be transformed into feasible beings \citep{fayard2022cap}.  
In a spatial accessibility setting, for instance, the capability set could be the set of all functionings reachable within a given travel budget.  
This flexibility allows the framework to encompass both discrete choice situations and continuous opportunity spaces.

Let \(\mathcal{C}\) be a compact subset of \(\Rhpos\), called the \emph{capability space}.  
We denote by boldface letters such as \(\A, \B, \C\) subsets of \(\mathcal{C}\).  
When these subsets are nonempty and compact, we call them \emph{capability sets}.  
Compactness expresses that feasible beings are both limited and well-defined: boundedness reflects the finiteness of available resources, and closure the idea that opportunity structures include all their attainable limits. Individuals are assumed to reason over attainable beings rather than over sequences that merely approach them.
We exclude the empty set, which would correspond to the absence of any viable being, in the Capability Approach, a situation equivalent to the loss of all functionings, \emph{i.e.}, death.
In some cases, it will be convenient to consider subsets of the larger space \(\Rhpos\),  
whenever we do so, we will make this explicit, and such sets are not referred to as capability sets.

For any subset \(\A \subseteq \mathcal{C}\), we define its \emph{Positive Domination Closure} (PDC) as
\begin{equation}\label{eq:PDC}
 \A^{\dm} = \{\vec{x}\in \Rhnneg \mid \exists \vec{a}\in \A \text{ such that } \vec{a}\geq\vec{x}\}.
\end{equation}
That is, \(\A^{\dm}\) consists of all beings that are weakly dominated by at least one element of \(\A\).

\ssk We now extend the weak and strict dominance relations to subsets of \(\Rhpos\).

\begin{defn}\label{def:setDominance}
Let \(\A\) and \(\B\) be subsets of $\Rhnneg$.  

We say that \(\A\) \emph{weakly dominates} \(\B\), denoted \(\A \geq \B\), if for every being \(\vec{b} \in \B\), there exists a being \(\vec{a} \in \A\) such that \(\vec{a} \geq \vec{b}\).  

We say that \(\A\) \emph{strictly dominates} \(\B\), denoted \(\A > \B\), if \(\A \geq \B\) and there exists at least one being \(\vec{a} \in \A\) such that no being \(\vec{b} \in \B\) satisfies \(\vec{b} \geq \vec{a}\).  

\smallskip
We say that \(\A\) \emph{strongly dominates} \(\B\), denoted \(\A \gg \B\),  
if \(\A \geq \B\) and there exist a being \(\vec{a} \in \A\) and \(\varepsilon > 0\) such that  
the hyper-rectangle \([\vec{a} - \varepsilon \cdot \vec{1}, \vec{a}]\) is contained in \(\Rhpos\)  
and does not intersect the dominated region of \(\B\), that is,
\[
[\vec{a} - \varepsilon \cdot \vec{1}, \vec{a}] \cap \B^\dm = \emptyset,
\]
where \(\vec{1} = (1, \ldots, 1)\) denotes the unit vector in \(\mathbb{R}^{h^*}\)\footnote{%
The notation \([\vec{a} - \varepsilon \cdot \vec{1}, \vec{a}]\) refers to the closed hyper-rectangle generated by the componentwise intervals \([a_h - \varepsilon, a_h]\) for each dimension \(h\).}.
  
\end{defn}

If \(\A\) and \(\B\) are capability sets (i.e., nonempty compact subsets of \(\mathcal{C}\)),  
then \(\A \geq \B\) means that \(\B\) is contained in the dominated region of \(\A\), that is, \(\B \subseteq \A^\dm\).  
Similarly, \(\A > \B\) means that \(\B \subset \A^\dm\).  

\ssk The distinction between strict and strong dominance lies in the notion of a \emph{perceptible expansion of freedom}.  
While strict dominance (\(\A > \B\)) only requires that \(\A\) includes at least one being not dominated by any element of \(\B\),  
strong dominance (\(\A \gg \B\)) additionally requires that this advantage be measurable:  
\(\A\) must open a non-negligible region of feasible beings that remains out of reach for \(\B\).  
The existence of an \(\varepsilon\)-gap ensures that the improvement is not merely theoretical  
(for instance, adding an isolated irrational point among rationals)  
but corresponds to an actual, perceivable enlargement of the capability space.  
This interpretation is consistent with the Capability Approach,  
where differences in freedom should reflect tangible opportunities  
that agents can realistically experience or value.

\begin{ex}
Figure \ref{fig:Three_Capability_set} depicts three capability sets in the case $h^* = 2$. 
Two of the capability sets, denoted $\A$ and $\B$, are finite, while the third 
one, denoted $\C$, contains an infinite number of solutions. Areas in the plot represent the spaces dominated by $\A$, $\B$, and $\C$, specifically $\A^\dm$, $\B^\dm$, and $\C^\dm$.

In this example, $\A$ should be preferred over $\B$ because $\B^\dm$ is entirely contained within $\A^{\dm}$. On the other hand, when comparing $\C$ to $\A$ (and $\B$) using this simple method, no comparison can be made in principle. This is because neither $\C$ is a subset of $\A^\dm$, nor is $\A$ a subset of $\C^\dm$.\hfill $\triangle$ 
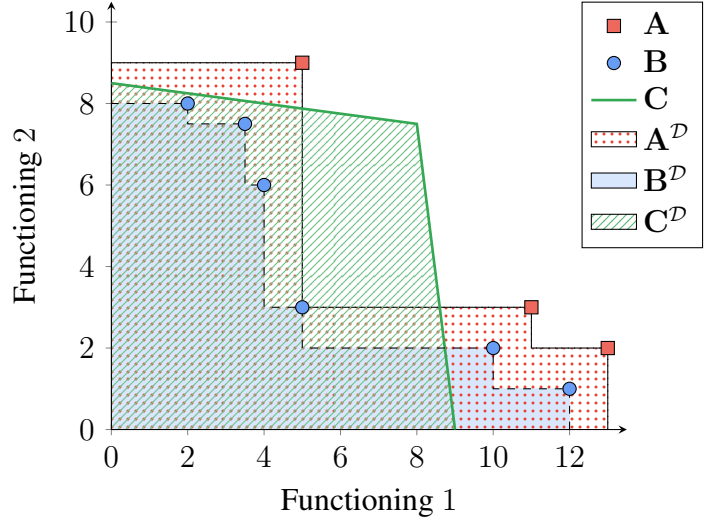
\begin{figure}
\centering
\begin{tikzpicture}
\begin{axis}[
    title={},
    xlabel={Functioning $1$},
    ylabel={Functioning $2$},
    xmin=0, xmax=13.5,
    ymin=0, ymax=10.5,
    grid=minor,
    axis x line=bottom, axis y line = left,
    legend style={at={(1.03,1)},anchor=north},
    legend cell align={left}]
    
\draw [blue!20,fill=blue!20] (0,0) --(12, 0)--(12, 1)--(10, 1)--(10, 2)--(5, 2)--(5, 3)--(4, 3)--(4, 6)--(3.5, 6)--(3.5, 7.5)--(2, 7.5)--(2, 8)--(0, 8) --(0,0)-- cycle;

\draw[pattern=dots, pattern color= red] (0,0) --(13, 0)--(13, 2)--(11, 2)--(11, 3)--(5, 3)--(5, 9)--(0, 9) --(0,0)-- cycle;

\draw[pattern=north east lines, pattern color=green!70] (0,0) -- (9,0) -- (8,7.5) -- (0,8.5) -- cycle;

\addplot[fill=red!80, mark = square*, only marks, mark size = 2.5pt]
coordinates {(13, 2)(11, 3)(5, 9)};
\addplot[fill=blue!80, mark = *, only marks, mark size = 2.5pt]
coordinates {(12, 1)(10, 2)(5, 3)(4, 6)(3.5, 7.5)(2, 8)};
\addplot[line width=1pt, color = green]
coordinates {(9, 0)(8, 7.5)(0, 8.5)};

\addlegendimage{area legend,pattern=dots, pattern color= red}
\addlegendimage{area legend,fill=blue!20}
\addlegendimage{area legend,pattern=north east lines, pattern color=green!80}
\addplot[dashed]
coordinates {(12, 0)(12, 1)(10, 1)(10, 2)(5, 2)(5, 3)(4, 3)(4, 6)(3.5, 6)(3.5, 7.5)(2, 7.5)(2, 8)(0, 8)};
\addplot[dotted]
coordinates {(13, 0)(13, 2)(11, 2)(11, 3)(5, 3)(5, 9)(0, 9)};

\addlegendentry{$\A$}
\addlegendentry{$\B$}
\addlegendentry{$\C$}
\addlegendentry{$\A^\dm$}
\addlegendentry{$\B^\dm$}
\addlegendentry{$\C^\dm$}
\end{axis}
\end{tikzpicture}
\caption{Three Capability sets and their PDC}
\label{fig:Three_Capability_set}
\end{figure}
\end{ex}

\ssk If \(\A\) is a capability set, the set of its non-strictly dominated points plays an important role in the sequel. 

\begin{defn}\label{def:Pareto}
Let \(\A \subseteq \Rhpos\) and \(\vec{a} \in \A\).  
The point \(\vec{a}\) is \emph{non-strictly dominated} in \(\A\) if there is no point \(\vec{b} \in \A\) such that \(\vec{b} > \vec{a}\).  
The set of all non-strictly dominated points of \(\A\), also called the \emph{Pareto frontier} of \(\A\), is denoted \(\nsdom[\A]\).
\end{defn}

In general, if we do not assume that \(\A\) is compact, its Pareto frontier \(\nsdom[\A]\) may be empty.  
If \(\A\) is a capability set (i.e., a nonempty compact subset of \(\mathcal{C}\)), then every point of \(\A\) is weakly dominated by a point in \(\nsdom[\A]\).

\begin{prop}\label{prop:compactPareto}
If \(\A \subseteq \Rhpos\) is compact, then for all \(\vec{a} \in \A\), there exists \(\vec{b} \in \nsdom[\A]\) such that \(\vec{b} \geq \vec{a}\).  
\end{prop}
\bpr
For any $\vec{x} \in \A$, let $\vec{x}^{\geq}$ denote the set of points in $\Rhnneg$ that weakly dominate $\vec{x}$, \ie $\vec{x}^{\geq} = \{\vec{y} \in \Rhnneg | \vec{y} \geq \vec{x}\}$. Consider the set $\A \cap \vec{x}^{\geq}$. It is a bounded and closed subset of $\Rhnneg$, since $\A$ is bounded and closed, and $\vec{x}^{\geq}$ is closed. Let $\vec{y}^* \in \arg\max\{\sum_{i=1}^{h^*} y_i, \vec{y} \in \A \cap \vec{x}^{\geq} \}$. We have $\vec{y}^* \in A \cap x^{\geq}$ since  the continuous function $\sum_{i=1}^{h^*} y_i$ on the compact set $\A \cap \vec{x}^{\geq}$ reaches its maximum in this set. The point $\vec{y}^*$ weakly dominates $\vec{x}$ and is non-strictly dominated in $\A \cap \vec{x}^{\geq}$. To prove the latter, assume by contradiction that there is $\vec{z}  \in \A \cap \vec{x}^{\geq}$ with $\vec{z} \neq \vec{y}$ and $\vec{z} > \vec{y}^*$. This would imply $\sum_{i=1}^{h^*} z_i > \sum_{i=1}^{h^*} y^*_i$, a contradiction. We just showed that $\vec{y}^*$ is non-strictly dominated in $\A \cap \vec{x}^{\geq}$. It is not either in $\A$ because no point in $\A \cap \vec{x}^{\geq}$ is dominated by a point in $\A \setminus \vec{x}^{\geq}$.  
\epr

An immediate consequence of this result is the following.
\begin{cor}
If $\A$ is a capability set, we have that $\A^{\dm} = \nsdom^{\dm}$. 
\end{cor}
\bpr
By Proposition~\ref{prop:compactPareto} and the transitivity of the weak dominance relation, any point weakly dominated by a point of $\A$ is weakly dominated by a point in $\nsdom$. 
\epr

\ssk Note that the Pareto frontier $P(\A)$ of a capability set $\A$ is not always a capability set. This is because the set $P(\A)$ may fail to be closed. 

\msk We consider a set of valuation functions \( v \in V \) such that each \( v: \mathcal{C} \to \mathbb{R}_{\geq0} \) maps the capability space \( \mathcal{C} \) to non-negative real numbers. The considered functions \( v(\cdot) \) are both increasing and continuous, the former meaning that:
$$
\forall v \in V, \forall (\vec{c}, \vec{c}^{\,\prime}) \in \mathcal{C}^2,\;\; \text{if } \vec{c} > \vec{c}^{\,\prime}, \text{ then } v(\vec{c}) > v(\vec{c}^{\,\prime}).
$$
The requirement for $v$ being defined on $\mathcal{C}$ is strong because it requires the ability to compare all the options in $\mathcal{C}$, which may not always be possible. 
Nonetheless, the monotonicity of \( v \) is a reasonable demand, given the objective of maximizing welfare across all dimensions.

\section{Freedom of choice}\label{Freedom}

As highlighted in the introduction, we want to consider two crucial aspects when comparing capability sets: their overall level of welfare (valuation of beings) and their distribution across different dimensions (diversity). 


One important aspect that is often overlooked in the measurement of capability sets \emph{is the value assigned to the freedom of choosing an option valued at a certain level of well-being}.
In the example given by \cite{sen1985commodities}, it may be reasonable to argue that a person who assigns no value to choice would be indifferent between a capability set where only $\vec{a}$ is achievable and a capability set where $\vec{a}$ is achievable along with another option $\vec{a^\prime}$ such that $v(\vec{a^\prime})<v(\vec{a})$.
However, it can also be argued that a person who values choice would prefer the second capability set over the first. 
This consideration highlights the importance of incorporating individual preferences over freedom and values over beings within the measurement framework.

The assessment of capability sets is a complex task that requires careful consideration. 
One way to do so is to count the number of options in the capability set, as proposed by \cite{pattanaik1990ranking} in an axiomatic characterization. 
However, it should be noted that the value of choice cannot be reduced to the cardinality of the capability set due to several important reasons:
\begin{itemize}

\item First, the existence of capability sets with an \emph{infinite number of beings}.
We need to be able to discern between different infinite sets in order to effectively evaluate their relative goodness. 
For instance, in Figure \ref{fig:ex1inf}, capability sets $\A$ and $\B$ are both infinite, yet it is evident that $\A$ dominates $\B$ in terms of the ordering of capabilities.
See \cite{pattanaik2000ranking} for a quantity-based  axiomatic work on infinite capability sets.
\captionsetup{margin=10pt, font=small, labelfont=bf}
\begin{figure}[ht]
    \centering
    \begin{minipage}[t]{.48\textwidth}
        \centering
        \resizebox{\columnwidth}{!}{%
\begin{tikzpicture}
\begin{axis}[
    xlabel={$h_1$},
    ylabel={$h_2$},
    xmin=0, xmax=8.5,
    ymin=0, ymax=8.5,
    grid=minor,
    axis x line=bottom,
    axis y line=left,
    legend style={at={(0.98,0.82)},anchor=south east,font=\small},
    legend cell align={left}
]

\draw[fill=blue!20, draw=black] 
  (0,0) -- (8,0) -- (0,7) -- cycle;

\draw[pattern=north west lines, pattern color=black!70, draw=black]
  (0,0) -- (4,0) -- (0,5) -- cycle;


\addlegendimage{area legend,fill=blue!20,draw=black}
\addlegendimage{area legend,pattern=north west lines, pattern color=black!70,draw=black}
\addlegendentry{$\A$}
\addlegendentry{$\B$}

\end{axis}
\end{tikzpicture}
}
    \caption{Comparison of Capability Sets $\A$ and $\B$ with infinite beings ($\A$ dominates $\B$)}
    \label{fig:ex1inf}
    \end{minipage}%
    \hspace{.04\textwidth}%
    \begin{minipage}[t]{.48\textwidth}
        \centering
        \resizebox{\columnwidth}{!}{%
\begin{tikzpicture}
\begin{axis}[
    xlabel={$h_1$},
    ylabel={$h_2$},
    xmin=0, xmax=8.5,
    ymin=0, ymax=8.5,
    grid=minor,
    axis x line=bottom,
    axis y line=left,
    legend style={at={(0.98,0.82)},anchor=south east,font=\small},
    legend cell align={left}
]

\addplot[
    only marks,
    mark=square*,
    mark size=2.5pt,
    fill=red!80
] coordinates {
(2,6.5) (2,6) (2.3,5.6) (1.8,6.8) (1.5,6.5)
(1.75,5.5) (1.7,5.8) (2.2,6.4) (1.9,6.4) (1.8,6.2)
};

\addplot[
    only marks,
    mark=*,
    mark size=2.5pt,
    fill=blue!80
] coordinates {
(1.7,6.4) (7.9,3.1)
};

\addlegendimage{only marks, mark=*, mark size=2.5pt, fill=blue!80}
\addlegendimage{only marks, mark=square*, mark size=2.5pt, fill=red!80}
\addlegendentry{$\A$}
\addlegendentry{$\B$}


\end{axis}
\end{tikzpicture}
}
        \caption{Comparison of Capability Sets $\A$ and $\B$. $\B$ has more potential beings, but $\A$ is more diverse}
    \label{fig:ex2} 
    \end{minipage}
\end{figure}

\begin{figure}[ht]
    \centering
    \begin{minipage}[t]{.48\textwidth}
        \centering
        \resizebox{\columnwidth}{!}{%
\begin{tikzpicture}
\begin{axis}[
    xlabel={$h_1$},
    ylabel={$h_2$},
    xmin=0, xmax=8.5,
    ymin=0, ymax=8.5,
    grid=minor,
    axis x line=bottom,
    axis y line=left,
    legend style={at={(0.98,0.82)},anchor=south east,font=\small},
    legend cell align={left}
]

\addplot[
    only marks,
    mark=square*,
    mark size=2.5pt,
    fill=red!80
] coordinates {
(1,6.4) (1.5,5) (3,3) (4,2.5)
(5,2) (6,1.7) (1.2,5.8) (7.6,1)
};

\addplot[
    only marks,
    mark=*,
    mark size=2.5pt,
    fill=blue!80
] coordinates {
(1.7,6.5) (8,3.1)
};


\addlegendimage{only marks, mark=*, mark size=2.5pt, fill=blue!80}
\addlegendimage{only marks, mark=square*, mark size=2.5pt, fill=red!80}
\addlegendentry{$\A$}
\addlegendentry{$\B$}

\end{axis}
\end{tikzpicture}
}

    \caption{Comparison of Capability Sets $\A$ and $\B$. $\B$ offers more options and is more diverse compared to $\A$, but every being of $\B$ is dominated by a being in $\A$}
    \label{fig:ex3}
    \end{minipage}%
    \hspace{.04\textwidth}%
    \begin{minipage}[t]{.48\textwidth}
        \centering
        \resizebox{\columnwidth}{!}{%
\begin{tikzpicture}
\begin{axis}[
    xlabel={$h_1$},
    ylabel={$h_2$},
    xmin=0, xmax=8.5,
    ymin=0, ymax=8.5,
    grid=minor,
    axis x line=bottom,
    axis y line=left,
    legend style={at={(0.98,0.82)},anchor=south east,font=\small},
    legend cell align={left}
]

\draw[fill=blue!20, draw=black]
  (0,0) -- (8,0) -- (8,3) -- (5,3) -- (5,5) -- (3,5) -- (3,7) -- (0,7) -- cycle;

\addplot[
    only marks,
    mark=square*,
    mark size=2.5pt,
    fill=red!80
] coordinates {
(7.95,2.95)
(4.95,4.95)
(2.95,6.95)
};

\addlegendimage{area legend,fill=blue!20,draw=black}
\addlegendimage{only marks,mark=square*,mark size=2.5pt,fill=red!80}
\addlegendentry{$\A$}
\addlegendentry{$\B$}


\end{axis}
\end{tikzpicture}
}
    \caption{Two equivalent sets $\A, \B$ according to Axiom \ref{prop:indif}}
    \label{fig:pareto}
    \end{minipage}
\end{figure}

\item The second reason is linked to the notion of \emph{diversity} \citep{pattanaik2000diversity} of options (or their range, \citealt{klemisch1993freedom}). 
Let a capability set $\B$ represent ten possible ways of walking to work, while a capability set $\A$ represents one way of walking and one way of cycling to work. 
In this case, although $\B$ offers a higher number of possible beings, it can be argued that the diversity of beings in $\A$ is greater in terms of freedom of choice, as  Figure \ref{fig:ex2} illustrates.

\item Thirdly, we need to take into account the \emph{value} of the offered choices according to individual preferences \citep{sen1990welfare,sen1991welfare}. 
An individual may possess a capability set $\B$ that contains numerous options, including diverse but low-valued ones, and may perceive as having fewer capabilities compared to a capability set $\A$ that offers fewer options but of higher value.
For instance in Figure \ref{fig:ex3}, $\B$ has more options and is ``more diverse'' than $\A$, but every being of $\B$ is dominated by a being in $\A$.   
\end{itemize}

Following these remarks, we assume that including dominated or redundant beings in the capability space does not enhance individual freedom. 
For instance, adding the option to ``go to work using only one leg'' when ``going to work on foot'' is already available does not improve an individual's freedom. 
Similarly, duplicating an existing opportunity, such as adding to $\A=\{\vec{a}\}$ an identical element $\vec{b}=\vec{a}$, does not expand the space of attainable beings. 
In both cases, the set of feasible functionings remains unchanged, as these additions do not modify the Pareto frontier. 
Hence, a capability set can be reduced to its Pareto frontier without any loss of relevant information (see Figure~\ref{fig:pareto}).

\msk 
We now introduce a function $\Phi$ that evaluates capability sets. 
Formally, for a capability set $\A \subseteq \mathcal{C}$, a value function $v \in V$, and a \emph{freedom-sensitivity function} $\phi : \mathbb{R}_{\ge 0} \to \mathbb{R}_{\ge 0}$, assumed to be continuous and strictly positive except perhaps in 0,  
we denote by \(\Phi_v^{\phi}(\A)\) the value assigned to \(\A\) according to \(v\), once transformed by \(\phi\).   
We denote by $\Phi_v^{\phi}(\A)$ the value assigned to $\A$ according to $v$, after transforming $v$ by $\phi$. 
This evaluation takes a non-negative real value: $\Phi_v^{\phi}(\A) \in \mathbb{R}_{\ge 0}$. 

\ssk 
Intuitively, the freedom-sensitivity function $\phi$ determines how the individual values differences in attainable well-being levels. 
It transforms the valuation $v(\vec{a})$ of each attainable being $\vec{a} \in \A$ into a modified contribution $\phi(v(\vec{a}))$, 
thereby shaping how freedom responds to the distribution of values within the capability set. 
A convex $\phi$ amplifies higher valuations of $v$, approximating an \emph{instrumental} view of freedom where the best options dominate the assessment. 
Conversely, a concave $\phi$ gives relatively greater importance to lower-valued beings, capturing an \emph{intrinsic} or egalitarian view of freedom. 

In view of the above discussion, we assume that $\Phi_v^{\phi}$ satisfies the following property.
\begin{ax}[Indifference of insignificant beings]\label{prop:indif}
For all value function $v\in V$, all freedom-sensitivity
function $\phi$ and all capability sets $\A, \B$ such that $\A \geq \B$, we have 
$\Phi^\phi_v(\A) = \Phi^\phi_v(\A\cup \B)$.
\end{ax} 

Axiom~\ref{prop:indif} entails the following. 
 
\begin{prop}\label{prop:ADPAcapaSets}
If $\A$ is a capability set, $\A^{\dm}$ is also a capability set. Under Axiom~\ref{prop:indif}, $\Phi^\phi_v(\A) = \Phi^\phi_v(\A^{\dm})$. In case $\nsdom$ is a capability set, we also have $\Phi^\phi_v(\A)= \Phi^\phi_v(\nsdom)$. \end{prop}

\bpr
Although it seems intuitively clear that $\A^{\dm}$ is compact as soon as $\A$ is compact, we could not find a proof of this result in literature. We thus provide a -- not straightforward -- proof in Appendix~\ref{app:proofADcapaSet} (Lemma~\ref{lem:proofADcapaSet}). 

\ssk Assuming that $\A$ is a capability set, we have $\A^{\dm} \geq \A$ and $\A \geq \A^{\dm}$. Applying Axiom~\ref{prop:indif}, we get $\Phi^\phi_v(\A) = \Phi^\phi_v(\A \cup \A^{\dm}) = \Phi^\phi_v(A^{\dm})$. 

\ssk Similarly, if $\nsdom$ is closed, hence a capability set, we have $\Phi^\phi_v(\nsdom) = \Phi^\phi_v(\A^{\dm}) = \Phi^\phi_v(\A)$, since $\nsdom^{\dm} = \A^{\dm}$. \epr

\ssk \cite{foster2011freedom} argues that a measure of freedom depends on the quantity and quality of available beings.  
However, we contend that the notion of diversity, rather than quantity, is a more appropriate indicator of freedom. Specifically, we propose that freedom would emerge from a \emph{conjoint measure of diversity and valuation of options}.
Hence, the concept of a Pareto dominant set, which focuses on the set of solutions that are not dominated by any other solution in the space, is perceived more suitable in this context.

\ssk For instance, consider options from sets $\A$ and $\B$, where $\A$ may have an infinite number of solutions while $\B$ has only three, as depicted in Figure \ref{fig:pareto}. Despite this difference, these solutions can be considered equivalent under certain circumstances (i.e., Axiom \ref{prop:indif}). This is because, if all relevant dimensions are accounted for, a rational individual endowed with the capability set $\A$ would invariably opt for a solution from $\B$. Therefore, a measure of freedom based on diversity, rather than quantity or quality alone, is more appropriate for capturing the essence of freedom in the capability approach.

\subsection{Gaertner and Xu method}

\ssk To the authors' knowledge, the only works that have tried to measure freedom as jointly involving the diversity and the valuation of options are \cite{gaertner2012evaluating,gaertner2006capability,gaertner2008new,gaertner2011reference}.
It is worth noting that these methods have been primarily proposed for assessing human development, whereas the current study addresses capabilities of all types.
The fundamental principles of their model are the following (with adaptations made to facilitate understanding in the context of our problem):

\ssk First, a reference beings vector $\vec{k^0}$ in the capability space $\mathcal{C}$ is defined, which represents the direction of societal development. 
Typically, this vector indicates a boundary of deprivation below which an individual is considered poor. Alternatively, it can be the existing average level of development.

Next, a distance function $d(x, y)$ between two points $x$ and $y$ is defined.\footnote{The distance function need to respect some properties (see \citealt{gaertner2006capability}).} 
For instance, the Euclidean distance can be used, given by $$d(x, y) =\sqrt{\sum^{h^*}_{h=1}(x_h-y_h)^2},$$ where $h^*$ denotes the number of dimensions in the capability space.

Then, in an adapted version of
the Gaertner-Xu method, a capability set $\A$ is assessed as follows:
\begin{enumerate}
    \item if $\vec{k^0} \not \in \A^\dm$, the assessment is a negative number; it is equal to minus the radius of the smallest sphere centered in $\vec{k^0}$ that intersects the set $\A^\dm$ in a point $\vec{a}$ dominated by $\vec{k^0}$;
    \item if $\vec{k^0} \in \A^\dm$, the assessment is a positive number; it is equal to the radius of the largest sphere centered in $\vec{k^0}$ satisfying the following condition: the part of this sphere contained in the positive orthant centered in $\vec{k^0}$ is included in $\A^\dm$.
\end{enumerate}
Mathematically, the assessment $r(\A, d)$ of $\A$ is computed as follows:

\small $r(\A, d) = \begin{cases} -\min_t 
\{t\in \mathbb{R}_+ :\{\vec{a}\in \mathbb{R}^{h^*}_{\geq0} : \vec{a}\leq \vec{k^0}, d(\vec{a},\vec{k^0})\leq t\}\cap \A^\dm\neq \emptyset\}\>\mbox{if} \> \vec{k^0}\notin \A^\dm,\\
\max_t 
\{t\in \mathbb{R}_+ :\{\vec{a}\in \mathbb{R}^{h^*}_{\geq0} : \vec{a}\geq \vec{k^0}, d(\vec{a},\vec{k^0}) \leq t\}\subseteq \A^\dm\}\> \mbox{if} \> \vec{k^0}\in \A^\dm.
\end{cases}$
\normalsize

\begin{ex}\label{ex:xu1}
Figure \ref{fig:gaertner} displays an example of the determination of the score of four capability sets using the adapted Gaertner-Xu method with $\vec{k^0}= (4, 4)$. 
We would obtain the order $r(\A, d)<r(\B, d)< r(\C, d) < r(\D, d)$
with $r(\A, d) = -2.5$, $r(\B, d) = -1.5$, $r(\C, d) = 1$, $r(\D, d) = 2$.$\hfill \triangle$
\end{ex}
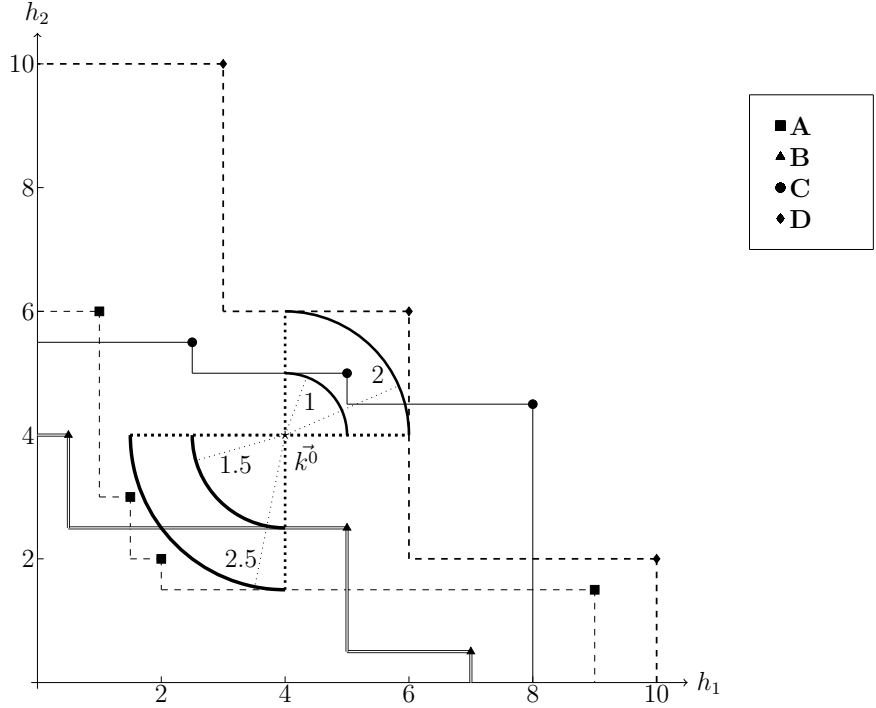
\begin{figure}
    \centering
\resizebox{0.7\textwidth}{!}{%
    \begin{tikzpicture}

    \draw[->] (-0.1,0) -- (10.5,0) node[right] {$h_1$};
    \draw[->] (0,-0.1) -- (0,10.5) node[above] {$h_2$};
    \draw[-] (2,0) -- (2,0.1) node[below] {$2$};
    \draw[-] (4,0) -- (4,0.1) node[below] {$4$};
    \draw[-] (6,0) -- (6,0.1) node[below] {$6$};
    \draw[-] (8,0) -- (8,0.1) node[below] {$8$};
    \draw[-] (10,0) -- (10,0.1) node[below] {$10$};
    \draw[-] (0,2) -- (0.1,2) node[left] {$2$};
    \draw[-] (0,4) -- (0.1,4) node[left] {$4$};
    \draw[-] (0,6) -- (0.1,6) node[left] {$6$};
    \draw[-] (0,8) -- (0.1,8) node[left] {$8$};
    \draw[-] (0,10) -- (0.1,10) node[left] {$10$};

\draw plot [only marks, mark=star] coordinates {(4,4)};

\draw [dashed] (0, 6) -- (1, 6);
\draw [dashed] (1, 6) -- (1, 3);
\draw [dashed] (1, 3) -- (1.5, 3);
\draw [dashed]  (1.5, 3) -- (1.5, 2);
\draw [dashed]  (1.5, 2) -- (2, 2);
\draw [dashed]  (2, 2) -- (2, 1.5);
\draw [dashed]  (2, 1.5) -- (9, 1.5);
\draw [dashed]  (9, 1.5) -- (9, 0);
\draw plot [only marks, mark=square*] coordinates {(1, 6)(1.5,3)(2,2)(9,1.5)};
\draw [double] (7, 0) -- (7, 0.5);
\draw [double]  (7, 0.5) -- (5, 0.5);
\draw [double]  (5, 0.5) -- (5, 2.5);
\draw [double]  (5, 2.5) -- (0.5, 2.5);
\draw [double]  (0.5, 2.5) -- (0.5, 4);
\draw [double]  (0.5, 4) -- (0, 4);
\draw plot [only marks, mark=triangle*] coordinates {(0.5, 4)(7,0.5)(5,2.5)};
\draw (8, 0) -- (8, 4.5);
\draw (8, 4.5) -- (5, 4.5);
\draw (5, 4.5) -- (5, 5);
\draw (5, 5) -- (2.5, 5);
\draw (2.5, 5) -- (2.5, 5.5);
\draw (2.5, 5.5) -- (0, 5.5);
\draw plot [only marks, mark=*] coordinates {(8,4.5)(5,5)(2.5,5.5)};
\draw [thick, dashed] (10, 0) -- (10, 2);
\draw [thick, dashed]  (10, 2) -- (6, 2);
\draw [thick, dashed]  (6, 2) -- (6, 6);
\draw [thick, dashed] (6, 6) -- (3, 6);
\draw [thick, dashed]  (3, 6) -- (3, 10);
\draw [thick, dashed]  (3, 10) -- (0, 10);
\draw plot [only marks, mark=diamond*] coordinates {(10,2)(6,6)(3,10)};
\draw [very thick, dotted] (1.5, 4) -- (6, 4);
\draw [very thick, dotted] (4-2.5, 4) -- (6, 4);
\draw [very thick, dotted] (4, 4-2.5) -- (4, 6);
\draw [ultra thick] (4,4-2.5) arc (-90:-180:2.5);
\draw [ultra thick] (4,2.5) arc (270:180:1.5);
\draw [very thick] (5,4) arc (0:90:1);
\draw [very thick] (6,4) arc (0:90:2);
\draw plot [only marks, mark = square*] (12,9);
\draw plot [only marks, mark = triangle*] (12,8.5);
\draw plot [only marks, mark = *] (12,8);
\draw plot [only marks, mark = diamond*] (12,7.5);
\draw (4,4) node[below right]{$\vec{k^0}$};
\draw (12,9) node[right]{$\A$};
\draw (12,8.5) node[right]{$\B$};
\draw (12,8) node[right]{$\C$};
\draw (12,7.5) node[right]{$\D$};
\draw (11.5, 9.5) -- (13.5, 9.5);
\draw (11.5, 7) -- (13.5, 7);
\draw (11.5, 9.5) -- (11.5, 7);
\draw (13.5, 9.5) -- (13.5, 7);
\draw [dotted] (4, 4) -- (5.84, 4.8);
\draw (5.5,4.7) node[above] {$2$};
\draw [dotted] (4, 4) -- (4.35, 4.94);
\draw (4.15, 4.54) node[right] {$1$};
\draw [dotted] (4, 4) -- (4-1.44, 4-0.41);
\draw (4 -0.8, 4 - 0.2) node[below] {$1.5$};
\draw [dotted] (4, 4) -- (4 -0.5, 4 - 2.5);
\draw (4 -0.3, 4 - 2) node[left] {$2.5$};
\end{tikzpicture}}
    \caption{An example of the determination of scores for some capability sets using Gaertner-Xu like method}
    \label{fig:gaertner}
\end{figure}

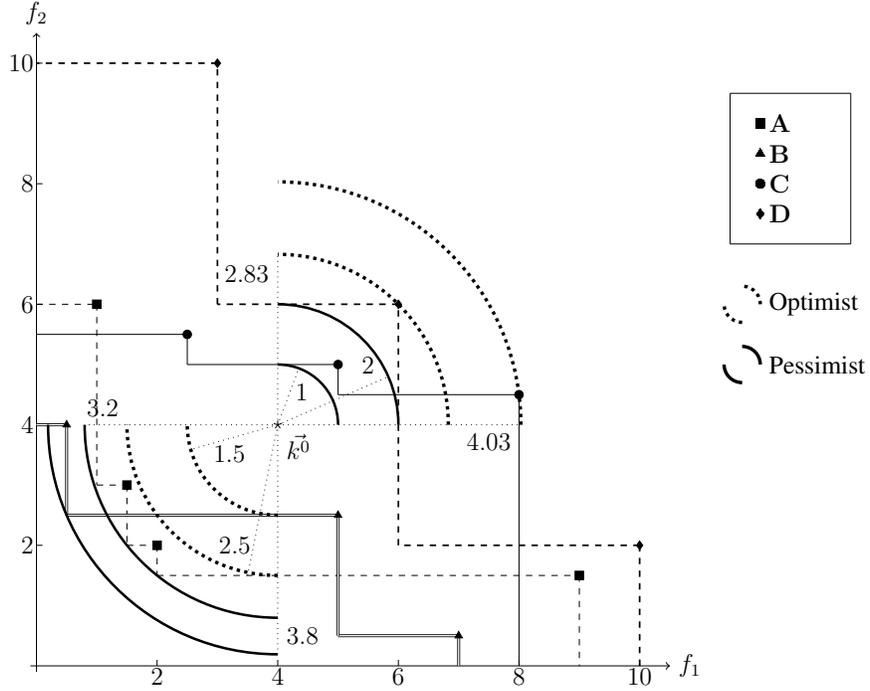
\begin{figure}
    \centering
    \resizebox{0.7\textwidth}{!}{%
    \begin{tikzpicture}

    \draw[->] (-0.1,0) -- (10.5,0) node[right] {$f_1$};
    \draw[->] (0,-0.1) -- (0,10.5) node[above] {$f_2$};
    \draw[-] (2,0) -- (2,0.1) node[below] {$2$};
    \draw[-] (4,0) -- (4,0.1) node[below] {$4$};
    \draw[-] (6,0) -- (6,0.1) node[below] {$6$};
    \draw[-] (8,0) -- (8,0.1) node[below] {$8$};
    \draw[-] (10,0) -- (10,0.1) node[below] {$10$};
    \draw[-] (0,2) -- (0.1,2) node[left] {$2$};
    \draw[-] (0,4) -- (0.1,4) node[left] {$4$};
    \draw[-] (0,6) -- (0.1,6) node[left] {$6$};
    \draw[-] (0,8) -- (0.1,8) node[left] {$8$};
    \draw[-] (0,10) -- (0.1,10) node[left] {$10$};
\draw plot [only marks, mark=star] coordinates {(4,4)};

\draw [dashed] (0, 6) -- (1, 6);
\draw [dashed] (1, 6) -- (1, 3);
\draw [dashed] (1, 3) -- (1.5, 3);
\draw [dashed]  (1.5, 3) -- (1.5, 2);
\draw [dashed]  (1.5, 2) -- (2, 2);
\draw [dashed]  (2, 2) -- (2, 1.5);
\draw [dashed]  (2, 1.5) -- (9, 1.5);
\draw [dashed]  (9, 1.5) -- (9, 0);
\draw plot [only marks, mark=square*] coordinates {(1, 6)(1.5,3)(2,2)(9,1.5)};

\draw [double] (7, 0) -- (7, 0.5);
\draw [double]  (7, 0.5) -- (5, 0.5);
\draw [double]  (5, 0.5) -- (5, 2.5);
\draw [double]  (5, 2.5) -- (0.5, 2.5);
\draw [double]  (0.5, 2.5) -- (0.5, 4);
\draw [double]  (0.5, 4) -- (0, 4);
\draw plot [only marks, mark=triangle*] coordinates {(0.5, 4)(7,0.5)(5,2.5)};

\draw (8, 0) -- (8, 4.5);
\draw (8, 4.5) -- (5, 4.5);
\draw (5, 4.5) -- (5, 5);
\draw (5, 5) -- (2.5, 5);
\draw (2.5, 5) -- (2.5, 5.5);
\draw (2.5, 5.5) -- (0, 5.5);
\draw plot [only marks, mark=*] coordinates {(8,4.5)(5,5)(2.5,5.5)};

\draw [thick, dashed] (10, 0) -- (10, 2);
\draw [thick, dashed]  (10, 2) -- (6, 2);
\draw [thick, dashed]  (6, 2) -- (6, 6);
\draw [thick, dashed] (6, 6) -- (3, 6);
\draw [thick, dashed]  (3, 6) -- (3, 10);
\draw [thick, dashed]  (3, 10) -- (0, 10);
\draw plot [only marks, mark=diamond*] coordinates {(10,2)(6,6)(3,10)};

\draw [ultra thick, dotted] (4,4-2.5) arc (-90:-180:2.5);
\draw [very thick] (4,4-3.807887) arc (-90:-180:3.807887);
\draw [ultra thick, dotted] (4,2.5) arc (270:180:1.5);
\draw [very thick] (4,4-3.201562) arc (-90:-180:3.201562);
\draw [very thick] (5,4) arc (0:90:1);
\draw [ultra thick, dotted] (4+4.031129,4) arc (0:90:4.031129);
\draw [very thick] (6,4) arc (0:90:2);
\draw [ultra thick, dotted] (4+2.828427
,4) arc (0:90:2.828427);

\draw plot [only marks, mark = square*] (12,9);
\draw plot [only marks, mark = triangle*] (12,8.5);
\draw plot [only marks, mark = *] (12,8);
\draw plot [only marks, mark = diamond*] (12,7.5);

\draw (4,4) node[below right]{$\vec{k^0}$};
\draw (12,9) node[right]{$\A$};
\draw (12,8.5) node[right]{$\B$};
\draw (12,8) node[right]{$\C$};
\draw (12,7.5) node[right]{$\D$};
\draw [ultra thick, dotted] (12,6) arc (0:90:0.3 );
\draw [ultra thick, dotted] (11.7,5.7) arc (270:180:0.3);
\draw [ultra thick] (12,5) arc (0:90:0.3 );
\draw [ultra thick] (11.7,4.7) arc (270:180:0.3);
\draw (12,6) node[right]{Optimist};
\draw (12,5) node[right]{Pessimist};
\draw (11.5, 9.5) -- (13.5, 9.5);
\draw (11.5, 7) -- (13.5, 7);
\draw (11.5, 9.5) -- (11.5, 7);
\draw (13.5, 9.5) -- (13.5, 7);

\draw [dotted] (4, 4) -- (5.84, 4.8);
\draw (5.5,4.7) node[above] {$2$};
\draw [dotted] (4, 4) -- (4.35, 4.94);
\draw (4.15, 4.54) node[right] {$1$};
\draw [dotted] (4, 4) -- (4-1.44, 4-0.41);
\draw (4 -0.8, 4 - 0.2) node[below] {$1.5$};
\draw [dotted] (4, 4) -- (4 -0.5, 4 - 2.5);
\draw (4 -0.3, 4 - 2) node[left] {$2.5$};

\draw [dotted] (4, 4) -- (4 + 4.03, 4);
\draw (7.5,4) node[below] {$4.03$};
\draw [dotted] (4, 4) -- (4 , 4+ 2.83);
\draw (4,6.5) node[left] {$2.83$};

\draw [dotted] (4, 4) -- (4 -3.2, 4);
\draw (1.1,4) node[above] {$3.2$};
\draw [dotted] (4, 4) -- (4 , 4 - 3.8);
\draw (4,0.5) node[right] {$3.8$};

\end{tikzpicture}}
    \caption{An example of the determination of scores for some capability sets using optimistic and pessimistic Gaertner-Xu like methods}
    \label{fig:optpes}
\end{figure}

Several generalizations of the method have been proposed in \cite{gaertner2008new,gaertner2011reference} and \cite{gaertner2012evaluating}. However, for the purpose of understanding the main idea, the simple unique point $(\vec{a}^0)$ formulation suffices.
It is important to highlight 
that the Gaertner-Xu method can be perceived as a pessimistic approach when $\vec{k^0}$ is in the capability set ($\vec{k^0}\in \A^\dm$), and an optimistic approach when it is located ``above'' it ($\vec{k^0}\in\mathcal{C}\backslash\A^\dm$). 
Specifically, if $\vec{k^0}$ is ``above'' the capability set, its score is determined by the closest point weakly dominated by $\A$ from $\vec{k^0}$, while other points weakly dominated by both $\A$ and $\vec{k^0}$ would have a lower score. 
Conversely, if $\vec{k^0}$ is contained within $\A^\dm$, the score is equal to the closest point on the frontier of weakly dominated solutions by $\A$, but it is possible for solutions that dominate $\vec{k^0}$ to exist at a greater distance.

It is worth mentioning that other rules can be formulated, although they may not necessarily satisfy all the axioms proposed by Gaertner and Xu. For instance, an optimistic version of the procedure would be:

\noindent \small $r(\A, d) = \begin{cases} -\min_t 
\{t\in \mathbb{R}_{\geq0} :\{\vec{a}\in \mathbb{R}^{h^*}_{\geq0} : \vec{a}\leq \vec{k^0}; d(\vec{a},\vec{k^0})\leq t\}\cap \A^\dm\neq \emptyset\}\>\;\;\; \mbox{if} \> \vec{k^0}\notin \A^\dm,\\
\max_t 
\{t\in \mathbb{R}_{\geq0} :\{\vec{a}\in \mathbb{R}^{h^*}_{\geq0} : \vec{a}\geq \vec{k^0}, d(\vec{a},\vec{k^0}) \leq t\}\cap \A^\dm \neq \emptyset\}\> \mbox{if} \> \vec{k^0}\in \A^\dm,
\end{cases}$

\normalsize \noindent whereas a pessimistic approach would be:\\
\small $r(\A, d) = \begin{cases} -\min_t 
\{t\in \mathbb{R}_{\geq0} :\{\vec{a}\in \mathbb{R}^{h^*}_{\geq0} : \vec{a}\leq \vec{k^0}, d(\vec{a},\vec{k^0}) = t\}\subseteq \A^\dm \}\>\>\;\;\; \mbox{if} \> \vec{k^0}\notin \A^\dm,\\
\max_t 
\{t\in \mathbb{R}_{\geq0} :\{\vec{a}\in \mathbb{R}^{h^*}_{\geq0} : \vec{a}\geq \vec{k^0}, d(\vec{a},\vec{k^0}) \leq t\}\subseteq \A^\dm \}\> \mbox{if} \> \vec{k^0}\in \A^\dm.
\end{cases}$

\normalsize
\begin{ex}\label{ex:xu2}
Using the same capability sets \(\A,\B,\C, \D\) and reference point $\vec{a}^0$ as in Example~\ref{ex:xu1}, we obtain different rankings under the optimistic and pessimistic procedures, as illustrated in Figure~\ref{fig:optpes}.  

Under the \emph{optimistic} procedure, the resulting order is:
\[
r(\C,d) > r(\D,d) > r(\B,d) > r(\A,d),
\]
with corresponding scores:
\[
r(\A,d) = -2.5,\quad r(\B,d) = -1.50,\quad r(\C,d) = 4.03,\quad r(\D,d) = 2.83.
\]

Under the \emph{pessimistic} procedure, the order is reversed:
\[
r(\D,d) > r(\C,d) > r(\A,d) > r(\B,d),
\]
with scores:
\[
r(\A,d) = -3.20,\quad r(\B,d) = -3.80,\quad r(\C,d) = 1.00,\quad r(\D,d) = 2.00.
\]
\hfill $\triangle$
\end{ex}

Importantly, note that in all cases, the entire distribution of the capabilities is not taken into account. 
Indeed, we only look at the extreme solutions; those closest or farthest from $\vec{k^0}$. 

\subsection{On ranking linear budget sets in terms
of freedom of choice}
Another interesting approach for assessing opportunity sets is the method proposed by \cite{xu2004ranking}. Their idea is to measure the volume of the set $\A^\dm$, denoted $\mathrm{vol}(\A^\dm)$.
It is important to note that the framework in \cite{xu2004ranking} respects strict monotonicity over the volume (in contrast to the approach of Gartner and Xu), and its volume-based concept is closely related to the method we propose in Section \ref{function}.

\begin{figure}
\centering
         \begin{tikzpicture}
\begin{axis}[
    title={},
    xlabel={Functioning $1$},
    ylabel={Functioning$2$},
    xmin=0, xmax=10.5,
    ymin=0, ymax=10.5,
    grid=minor,
    axis x line=bottom, axis y line = left,
    legend style={at={(1.03,1)},anchor=north},
    legend cell align={left}]
\draw [blue!20,fill=blue!20] (0,0)--(3, 0)--(3, 7)--(2, 7)--(2, 8)--(0, 8) --(0,0)-- cycle;
 \draw[pattern=dots, pattern color= red] (0,0)--(0, 3)--(7, 3)--(7, 2)--(8, 2)--(8, 0) --(0,0)-- cycle;
\addplot[fill=red!80, mark = square*, only marks, mark size = 2.5pt]
coordinates {(7, 3)(8, 2)};
\addplot[fill=blue!80, mark = *, only marks, mark size = 2.5pt]
coordinates {(3, 7)(2, 8)};

\addlegendimage{area legend,pattern=dots, pattern color= red}
\addlegendimage{area legend,fill=blue!20}
\addplot[dashed]
coordinates {(3, 0)(3, 7)(2, 7)(2, 8)(0, 8)};

\addlegendentry{$\A$}.
\addlegendentry{$\B$};
\addlegendentry{$\A^\dm$};
\addlegendentry{$\B^\dm$};
\end{axis}
\end{tikzpicture}
        \caption{Two symmetric capability sets}
        \label{fig:symetry}
    \end{figure}
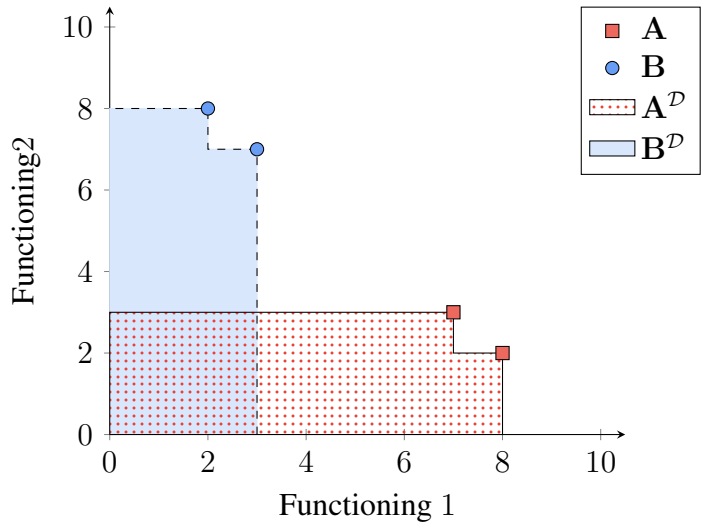

However, this method does not account for differences in the relative importance of welfare dimensions among citizens. This limitation becomes apparent through the fact that it satisfies the Symmetry Axiom, which states that for all $\A^\dm, \B^\dm, \C^\dm \in \mathcal{C}$, if $\A^\dm$ and $\B^\dm$ are symmetric, then
\[
\mathrm{vol}(\A^\dm)\geq \mathrm{vol}(\C^\dm) 
\;\;\Longleftrightarrow\;\; 
\mathrm{vol}(\B^\dm) \geq \mathrm{vol}(\C^\dm).
\]
While this axiom ensures that permutations of functiongs do not affect the ranking, it may be overly restrictive when individuals assign different importance to the functiongs. For instance, as illustrated in Figure \ref{fig:symetry}, if a person places greater value on functioning 1 relative to functioning 2, they would typically prefer set $\A$ over $\B$ rather than considering them equally desirable. By treating all welfare dimensions as equally important, the symmetry axiom overlooks such preference asymmetries, which can lead to less accurate rankings in contexts where the importance of functionings varies.

\section{Extremes of freedom valuation and desirable properties}\label{axio}
We begin by presenting two conceptual extremes in the evaluation of freedom. The first, which we refer to as the \emph{instrumental extreme} \citep{xu2003functionings,xu2003ranking}, assesses a capability set only based on its best-valued element, reflecting a view of freedom grounded in outcomes. In contrast, the \emph{intrinsic extreme} values freedom for its own sake, and considers that all equally desirable beings be accessible to achieve a given level of well-being. These two extremes define the normative upper and lower bounds within which our proposed compromise measure of freedom is situated (see Section \ref{function}).

\subsection{Instrumental extreme}

When individuals view freedom as having a purely \textit{instrumental value}, their concern is limited to the most favorable achievable outcome. In this perspective, the individual evaluates a capability set $\A$ solely based on the best element it contains. Consequently, they are indifferent between a set that contains only the best-valued solution and one that contains that solution along with additional, less desirable beings. The diversity or variety of options in the set does not matter; only the top performer does.

This behavior is captured by the following valuation function:
\[
\Phi^{\max}_v(\A) = \max_{\vec{a} \in \A^\dm} v(\vec{a}).
\]

We can associate to each capability set $\A$ a corresponding \textit{level set} of the valuation function $v$, which collects all elements in the capability space $\mathcal{C}$ that achieve the same value as the best element of $\A$:
\[
\A_{\max} = \left\{ \vec{c} \in \mathcal{C} \mid v(\vec{c}) = \Phi^{\max}_v(\A) \right\}.
\]
This level set captures the frontier of best-valued beings, according to the max criterion. 
It can be interpreted as a capability threshold: any capability set $\B$ such that
\(\Phi^{\mathrm{max}}_v(\B) \leq \Phi^{\mathrm{max}}_v(\A_{\mathrm{max}})\)
will be contained in $\A_{\mathrm{max}}^\dm$.

\begin{figure}
    \centering
    \includegraphics[width=0.55\linewidth]{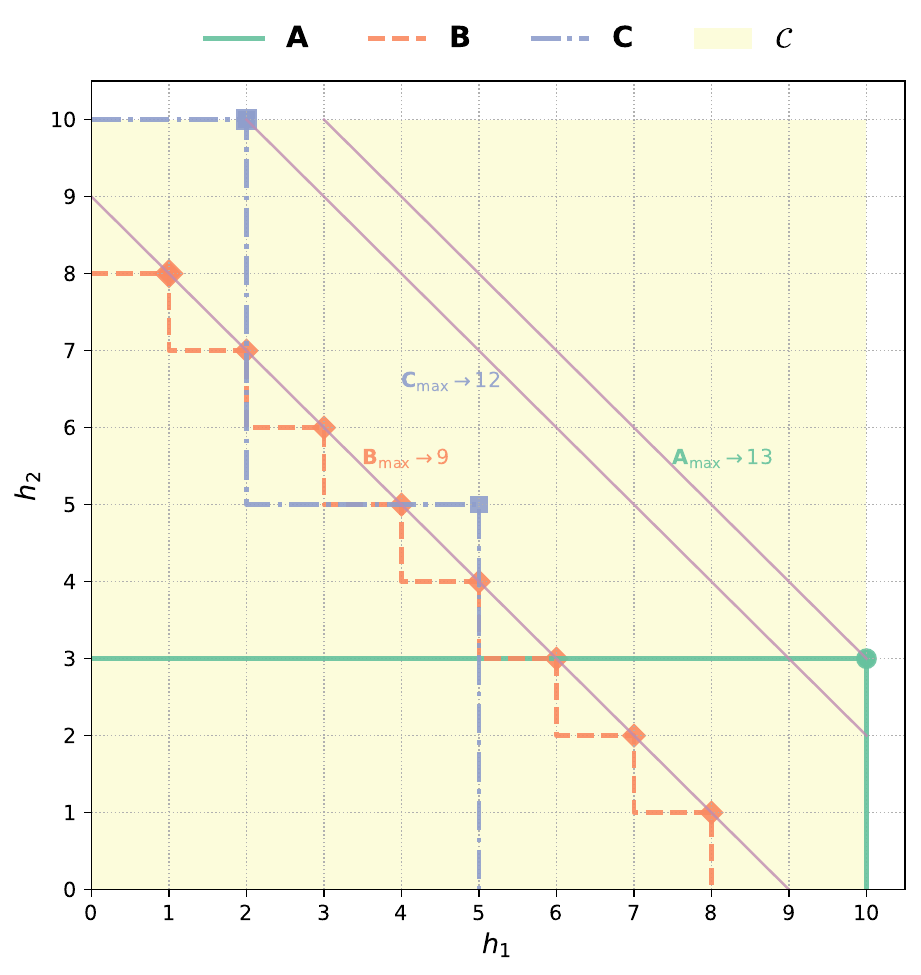}
\caption{Example 
\ref{ex:max_ass}: Instrumental extreme approach
    \label{fig:exampleG}}
\end{figure}

\begin{ex}\label{ex:max_ass}
Consider the following capability sets: \\
$\A = \{(10, 3)\}$, \\
$\B = \{(1, 8), (2, 7), (3, 6), (4, 5), (5, 4), (6, 3), (7, 2), (8, 1)\}$, \\
$\C = \{(2, 10), (5, 5)\}$,\\
together with the valuation function $v(\vec{c}) = c_1 + c_2$ defined over $\mathcal{C} = ((10, 10)^\dm) \cap \mathbb{R}^2_{\geq 0}$. This setup is illustrated in Figure \ref{fig:exampleG}.

In this case, $v(\vec{c})$ is a linear function, and the indifference curves are straight lines. Using the max approach, we obtain:
\[
\Phi^{\max}_v(\A) = 13, \quad 
\Phi^{\max}_v(\B) = 9, \quad 
\Phi^{\max}_v(\C) = 12.
\]
Each set is thus evaluated based solely on its best element under the valuation function. The corresponding level sets \( \A_{\max} \), \( \B_{\max} \), and \( \C_{\max} \), representing the sets of all beings in \( \mathcal{C} \) that share the same maximal value as each set, are also depicted in Figure~\ref{fig:exampleG}. \hfill $\triangle$
\end{ex}

Finally, the instrumental extreme approach can be seen as an ``optimistic'' interpretation of the Gaertner–Xu method. Specifically, when we take $\vec{k^0} = \vec{0}$ and define the distance function as $d(\vec{0}, \vec{a}) = v(\vec{a})$, we recover the max-based valuation.

\subsection{Intrinsic extreme}

An individual who values \emph{diversity} as an essential aspect of freedom may adopt a different perspective. From this point of view, a person considers themselves free to achieve a well-being level \( y \) only if they can access \emph{all beings} \( \vec{c} \in \mathcal{C} \) such that \( v(\vec{c}) = y \), or beings that are \emph{objectively better} in the sense that \( \vec{c^\prime} \geq \vec{c} \) and \( v(\vec{c}) = y \).  

In other words, they do not consider themselves free to reach level \( y \) unless their capability set includes all relevant solutions that attain (or surpass) that value.

This behavior is captured by the following valuation function:
\[
\Phi^{\min}_v(\A) = \inf_{\vec{c} \in \mathcal{C} \setminus \A^\dm} v(\vec{c}),
\]
which corresponds to the lowest well-being level that \emph{cannot} be achieved by the capability set \( \A \).

As with the extreme instrumental approach, we can define a level set representing all beings in \( \mathcal{C} \) that attain the boundary value of \( \Phi^{\min}_v(\A) \):
\[
\A_{\min} = \left\{ \vec{c} \in \mathcal{C} \mid v(\vec{c}) = \Phi^{\min}_v(\A) \right\}.
\]
This set captures the threshold beyond which the individual no longer considers themselves free to attain that level of well-being. 
For all capability sets $\B$, if we have $\Phi^{\min}_v(\A_{\min}) \leq \Phi^{\min}_v(\B)$, then $\A_{\min}$ is a set contained in $\B^\dm$.

\begin{figure}
    \centering\includegraphics[width=0.55\textwidth]{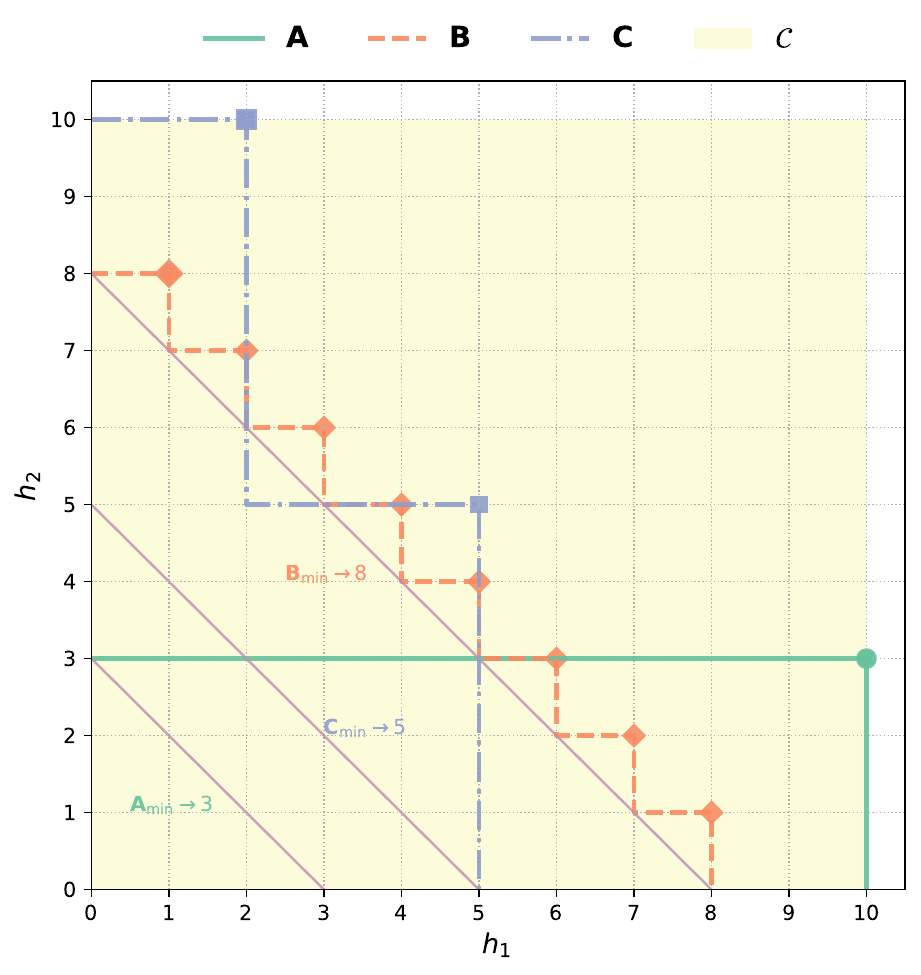}
    \caption{Example 
    \ref{ex:max_ass}: Intrinsic extreme approach}
    \label{fig:EXMIN}
\end{figure}

\addtocounter{ex}{-1}
\begin{ex}[Cont.]
Continuing from Example~\ref{ex:max_ass}, and using the same valuation function \( v(\vec{c}) = c_1 + c_2 \), we now apply the intrinsic extreme approach. As shown in Figure~\ref{fig:EXMIN}, we obtain:
\[
\Phi^{\min}_v(\A) = 3, \quad 
\Phi^{\min}_v(\B) = 7, \quad 
\Phi^{\min}_v(\C) = 5.
\]
This reflects the idea that \( \A \), while offering a high-value option, does not allow access to many other beings at similar or slightly lower levels of well-being. In contrast, \( \B \) offers a more uniform spread, covering a broader band of high-valued beings. The level sets \( \A_{\min}, \B_{\min}, \C_{\min} \) are illustrated in Figure~\ref{fig:EXMIN}. \hfill\(\triangle\)
\end{ex}

Finally, the intrinsic extreme approach corresponds to the ``pessimistic'' interpretation of the Gaertner–Xu method. This interpretation arises when we set \( \vec{a}^0 = \vec{0} \) and define the distance function as \( d(\vec{0}, \vec{b}) = v(\vec{b}) \), reflecting the minimal level not guaranteed by the capability set.

\subsection{Desirable properties}\label{ssec:desirable}

To ensure that our proposed freedom measure $\Phi$ is both theoretically sound and aligned with our intuitive understanding of freedom, it must satisfy a number of desirable properties. In the following, we outline a set of axioms that formalize these requirements over the capability sets and the valuation functions in $V$.

\begin{ax}[Strong Monotonicity] For all $v\in V$, all freedom-sensitivity
function $\phi$, and capability sets $\A, \B$, if $\A\geq \B$ then $\Phi^\phi_v(\A)\geq \Phi^\phi_v(\B)$, and if $\A \gg \B$ then $\Phi^\phi_v(\A) > \Phi^\phi_v(\B)$.
\end{ax}  
    \begin{ax}[Continuity] For all $v\in V$all freedom-sensitivity
function $\phi$, and capability sets $\A, \C$, such that $\Phi^\phi_v(\A) > \Phi^\phi_v(\C)$, for all \(\lambda \in [\Phi^\phi_v(\C), \Phi^\phi_v(\A)] \), there exists a capability set $\B$ such that $\A^\dm\cap\C^\dm\subseteq\B^\dm\subseteq\A^\dm\cup\C^\dm$ and $\Phi^\phi_v(\B)=\lambda$.
    \end{ax} 
        \begin{ax}[Invariance of Scaling Effects] For all $v\in V$, all freedom-sensitivity
function $\phi$, and capability sets $\A, \B$, if $\Phi^\phi_v(\A)\geq \Phi^{\phi}_v(\B)$ then, for any $\vec{\alpha}\in \mathbb{R}^{h^*}_{>0}$, letting $\A^\prime = \vec{\alpha}\cdot\A$, $\B^\prime = \vec{\alpha}\cdot\B$, and defining $v^\prime(\vec{\alpha}\cdot\vec{a})= v(\vec{a})$, we have
    \[
    \Phi^\phi_{v^\prime}(\A^\prime)\geq \Phi^\phi_{v^\prime}(\B^\prime).
    \]
    \end{ax} \begin{ax}[Bounded Freedom Principle] For all \( v \in V \), all freedom-sensitivity
function $\phi$, and for any capability set \( \A\),
    \[
    \Phi^\phi_v(\A_{\min}) \leq \Phi^\phi_v(\A) \leq \Phi^\phi_v(\A_{\max}).
    \]\end{ax} 

The \textbf{Strong Monotonicity} axiom requires that the valuation function \(\Phi_v\) respects the dominance relationships between capability sets in a way that reflects meaningful expansions of freedom.  
If one set \(\A\) weakly dominates another set \(\B\) (\(\A \geq \B\)), then its valuation should not be lower.  
If \(\A\) strongly dominates \(\B\) (\(\A \gg \B\)), then its valuation must be strictly higher.  
This axiom captures the fundamental intuition that freedom should increase whenever a capability set expands in a tangible and measurable manner, rather than by the mere addition of limit or redundant options.  
Variations of this principle, such as strict monotonicity \citep{xu2004ranking} and other forms of monotonicity \citep{kreps1979preference,puppe1996axiomatic,carter1995measure}, are common in the literature.  

Variations of this principle, such as strict monotonicity \citep{xu2004ranking} and other forms of monotonicity \citep{kreps1979preference,puppe1996axiomatic,carter1995measure}, are common in the literature.  

It is worth noting that the volume-based method of \cite{xu2004ranking} is defined for linear budget sets, where the dominance relation \(\A > \B\) directly implies \(\A \gg \B\).  
In the more general non-linear setting considered here, this implication no longer holds: \(\A > \B\) does not necessarily entail that \(\mathrm{vol}(\A^\dm) > \mathrm{vol}(\B^\dm)\).  
For instance, consider \(\A = \{(1,1)\}\) and \(\B = \{(1,1), (2,0)\}\).  
Here, \(\A < \B\) in the dominance sense, yet \(\mathrm{vol}(\A^\dm) = \mathrm{vol}(\B^\dm)\).

The \textbf{Continuity} axiom ensures that the valuation function \( \Phi_v \) behaves in a smooth and predictable manner when capability sets are modified. It reflects the intuition that freedom assessment should not change in a discontinuous or abrupt manner when capability sets evolve gradually.
In particular, this condition implies a stronger version of the \emph{Solvability} property, 
a standard requirement in Decision Analysis and measurement theory 
\citep{krantzLuceSuppesTversky1971}. A weaker variant of this axiom appears in \cite{gaertner2008new} under the name \emph{Betweenness}.

The \textbf{Invariance of Scaling Effects} axiom guarantees that the ordering of capability sets is invariant under uniform scaling of the underlying dimensions. That is, if every welfare dimension measure of a capability set is multiplied by a positive factor, the relative ranking of the freedom measures remains unchanged. This property is adapted from \cite{xu2004ranking} and is vital because the absolute levels of capabilities can depend on the units of measurement; however, the perceived degree of freedom should be independent of such scaling. Without this invariance, the freedom measure could be unduly influenced by arbitrary choices of scale, detracting from its consistency and applicability.

Finally, the \textbf{Bounded Freedom Principle} establishes that the freedom measure $\Phi^{\phi}_v(\A)$ is confined between two extremes: an intrinsic (min) and an instrumental (max) view of freedom. Specifically, it requires that
\[
\Phi^\phi_v(\A_{\min}) \leq \Phi^\phi_v(\A) \leq \Phi^\phi_v(\A_{\max})
\]
for every capability set $\A$, every $v\in V$ and every $\phi$. The intrinsic extreme represents the lower bound, reflecting a perspective that values beings for their inherent worth, while the instrumental extreme represents the upper bound, reducing the set to its best option. This axiom ensures that our measure of freedom remains within an interval that is rationally motivated .

Together, these axioms form the core properties that any reasonable measure of freedom should satisfy, thereby ensuring both theoretical rigor and practical relevance. 

\section{Compromise approach}\label{function}
In the previous approaches (instrumental and intrinsic), the evaluation of a capability set \(\A\) did not account for the contribution of all efficient solutions in its Pareto frontier.  
To overcome this limitation, we introduce a \emph{compromise approach} that integrates the contribution of every point in the Pareto frontier, ensuring that any addition of a non-dominated option expanding the measurable domain of attainable beings has a strictly positive impact on 
\(\Phi^\phi_v(\A)\).  
This approach satisfies the previously stated axioms, remains sensitive to the chosen value function \(v\), accommodates different perspectives on freedom, and offers a balanced and reasonable way to capture the value of capability sets.

\subsection{Definition}


We define the compromise method as follows:

\begin{gather*} 
    \Phi^\phi_v(\A) = \int\limits_{ \vec{a}\in\A^\dm}  \phi(v(\vec{a})) \,d\vec{a}
\end{gather*}
where $\phi$ is a continuous function $\phi: [0, \max_{\vec{a} \in \mathcal{C}} v(\vec{a})] \rightarrow \mathbb{R}_{\geq 0}$, with $\phi$ strictly positive except perhaps in $0$.

Proofs that $\Phi^\phi_v$ satisfies the desirable properties listed in Section~\ref{ssec:desirable} are presented in Appendix~\ref{app:proofsIntegralMethod}. 

\begin{prop}\label{prop:eqVol}
Let \(\phi: [0, \max_{\vec{a} \in \mathcal{C}} v(\vec{a})] \rightarrow \mathbb{R}_{\geq 0}\) be a constant function, i.e., \(\phi(v(\vec{a})) = c\) for all \(\vec{a} \in \A\). Then:
\[
\Phi^\phi_v(\A) = \int_{\vec{a} \in \A^\dm} \phi(v(\vec{a})) \, d\vec{a} = c \cdot \mathrm{vol}(\A^\dm).
\]
\end{prop}
See proof in Appendix \ref{app:eqVol}.

Several functions $\phi$ can be chosen to represent different behaviors towards freedom. 
Let us give three examples that represent different behaviors. 

\begin{figure}\centering
\includegraphics[width=0.7\textwidth]{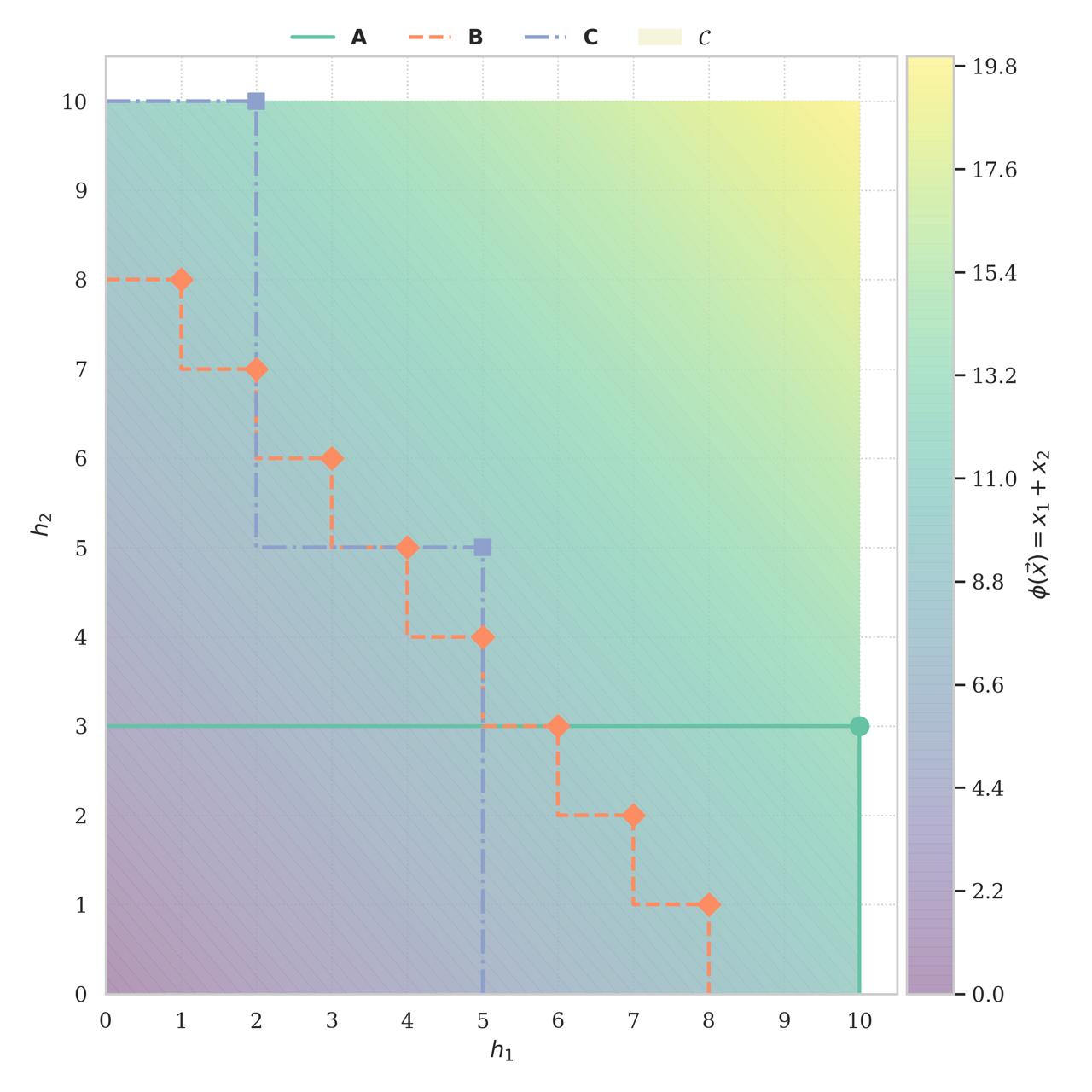}
    \caption{Example \ref{ex:max_ass}: $\Phi^\phi_v$ using the compromise approach with $\phi(v(\vec{a})) = v(\vec{a})$}
    \label{fig:alpha1}
\end{figure}

First, the individual considers the simple case where $\phi(v(\vec{a}))=v(\vec{a})$.
By applying this function to Example \ref{ex:max_ass}, we obtain the integration on the capability sets as in Figure \ref{fig:alpha1}. We obtain $\Phi^\phi_v(\A)=195$, $\Phi^\phi_v(\B)=204$ and $\Phi^\phi_v(\C)=210$. For instance to compute $\Phi^\phi_v(\A)$, we have to calculate the double integral over $(a_1, a_2) \in \A^\dm$, being $a_1 \in [0, 10]$ and $ a_2 \in [0, 3]$:
\[
\Phi^\phi_v(\A) = \iint\limits_{\vec{a}\in\A^\dm}= \phi(v(a_1,a_2))da_1da_2=\int_0^{10} \int_0^{3} (a_1 + a_2) \, da_2 \, da_1
\]
\[
= \int_0^{10} \left[ a_1a_2 + \frac{1}{2}a_2^2 \right]_0^3 \, da_1 = \int_0^{10} \left( 3a_1 + \frac{9}{2} \right) da_1
\]
\[
= \left[ \frac{3}{2}a_1^2 + \frac{9}{2}a_1 \right]_0^{10} = \frac{3}{2}(10)^2 + \frac{9}{2}(10) = 150 + 45 = 195.\]

\begin{figure}
\centering
    \includegraphics[width=0.7\textwidth]{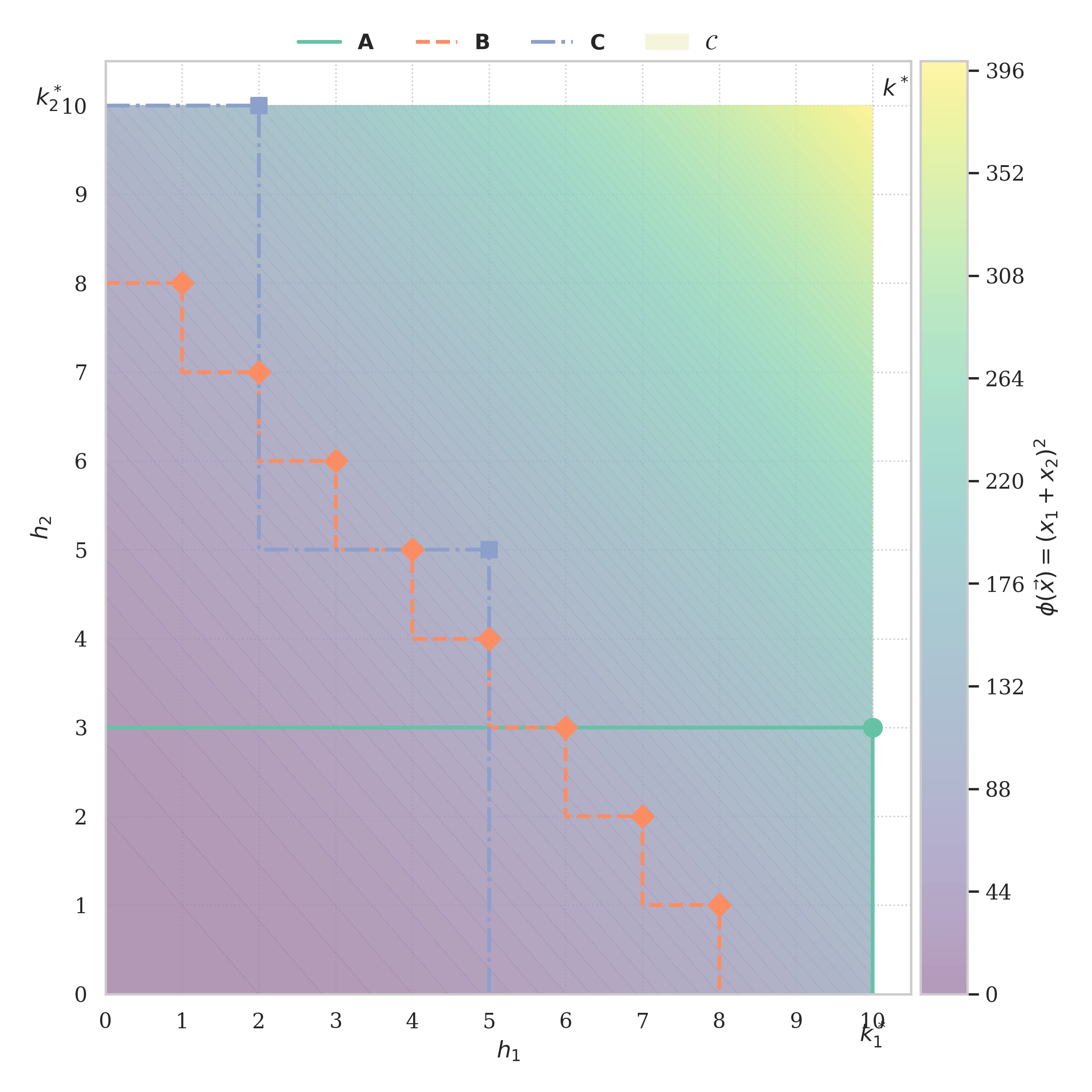}
    \caption{Example \ref{ex:max_ass}: $\Phi^\phi_v$ using the compromise approach with $\phi(v(\vec{a})) = v(\vec{a})^2$}
    \label{fig:alpha2}
\end{figure}

Secondly, the individual chooses a function \( \phi(v(\vec{a})) = v(\vec{a})^2 \), which amplifies the contribution of higher-valued options. In other words, beings associated with greater well-being receive disproportionately higher weight in the evaluation.
Continuing with Example \ref{ex:max_ass}, a new freedom valued capability set is displayed in Figure \ref{fig:alpha2}, and $\Phi^\phi_v(\A)=1,540$, $\Phi^\phi_v(\B)=1,302$ and $\Phi^\phi_v(\C)=1,475.8$. 
To compute $\Phi^\phi_v(\A)$, for instance, we do as follows:
\[
\Phi^\phi_v(\A) = \iint\limits_{\vec{a}\in\A^\dm} \phi(v(a_1,a_2)) \, da_1 \, da_2 = \int_0^{10} \int_0^{3} (a_1 + a_2)^2 \, da_2 \, da_1
\]
\[
= \int_0^{10} \left[ \frac{1}{3}(a_1 + a_2)^3 \right]_0^3 \, da_1 = \int_0^{10} \left( \frac{1}{3}(a_1 + 3)^3 - \frac{1}{3}a_1^3 \right) da_1
\]
\[
= \left[ \frac{1}{12}a_1^4 + a_1^3 + \frac{9}{2}a_1^2 \right]_0^{10} = \frac{1}{12}(10)^4 + (10)^3 + \frac{9}{2}(10)^2 = 1540.
\]

\begin{figure}
    \centering
    \includegraphics[width=0.7\textwidth]{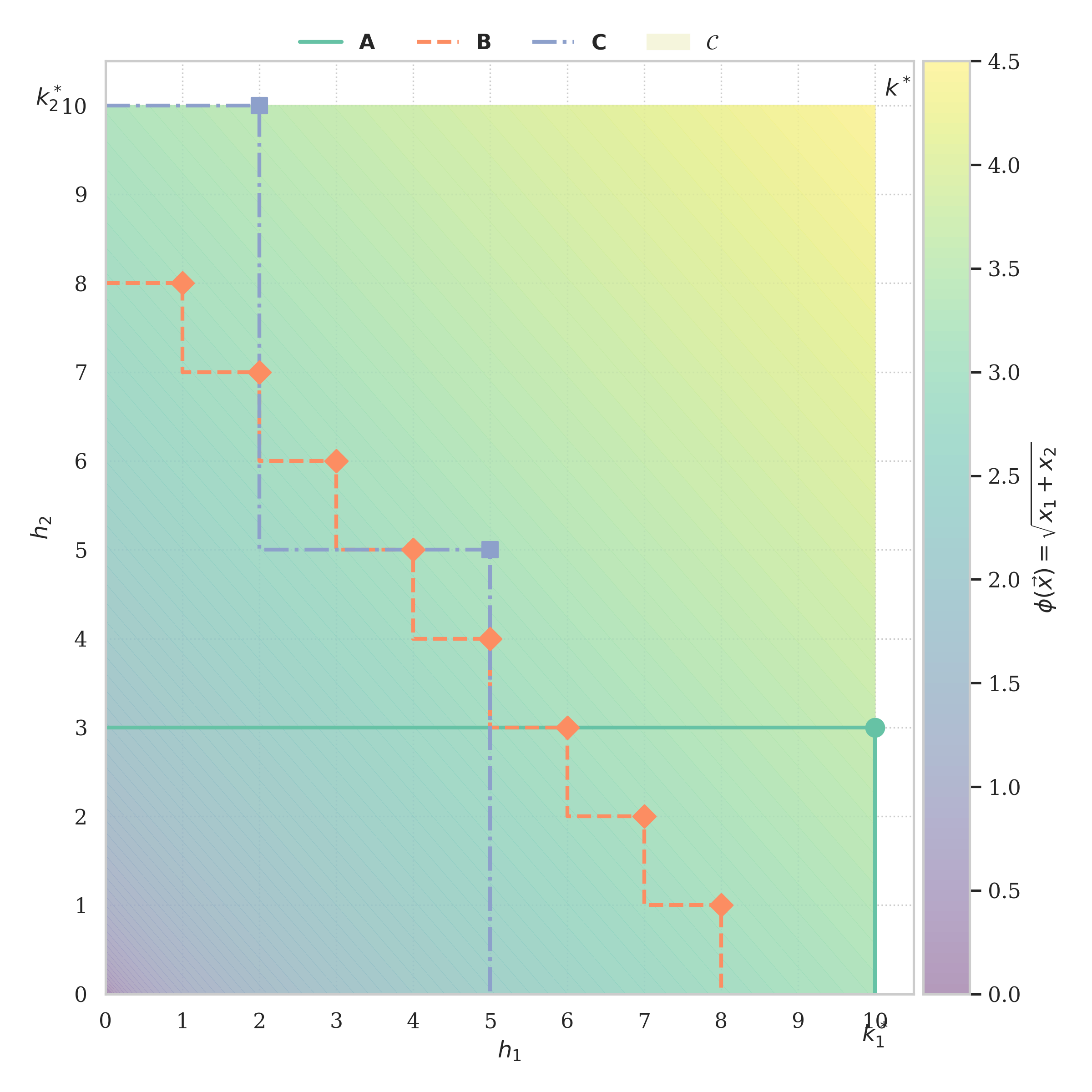}
\caption{Example \ref{ex:max_ass}: $\Phi^\phi_v$ using the compromise approach with $\phi(v(\vec{a})) = \sqrt{v(\vec{a})}$}
    \label{fig:alpha3}
\end{figure}

Thirdly, the individual's perspective prioritizes more balanced capability sets by assigning greater weight to lower-valued beings. An illustrative approach to achieve this balance is by applying the function \( \phi(v(\vec{a})) = \sqrt{v(\vec{a})} \).
Applying this method to our example, we compute the values \( \Phi^\phi_v(\A), \Phi^\phi_v(\B), \) and \( \Phi^\phi_v(\C) \) as depicted in Figure \ref{fig:alpha3}. The resulting calculations yield \( \Phi^\phi_v(\A) \approx 74.03 \), \( \Phi^\phi_v(\B) \approx 83.95 \), and \( \Phi^\phi_v(\C) \approx 83.55 \).
The computation of $\Phi^\phi_v(\A)$ is obtained as the following:
\[
\Phi^\phi_v(\A) = \iint\limits_{\vec{a}\in\A^\dm} \phi(v(a_1,a_2)) \, da_1 \, da_2 = \int_0^{10} \int_0^{3} \sqrt{a_1 + a_2} \, da_2 \, da_1
\]
\[
= \int_0^{10} \left[ \frac{2}{3}(a_1 + a_2)^{\frac{3}{2}} \right]_0^3 \, da_1= \int_0^{10} \left( \frac{2}{3}(a_1 + 3)^{\frac{3}{2}} - \frac{2}{3}a_1^{\frac{3}{2}} \right) da_1
\]
\[
=
\left[\dfrac{4\left(\left(a_1+3\right)^\frac{5}{2}-a_1^\frac{5}{2}\right)}{15}\right]_0^{10}
=
\dfrac{4\left(13^\frac{5}{2}-10^\frac{5}{2}-3^\frac{5}{2}\right)}{15}
 \approx 74.03.
\]

We note that by altering the function \( \phi \), we can emphasize different aspects of freedom. For example, choosing \( \phi(v(\vec{a})) = v(\vec{a})^2 \) accentuates the instrumental value of freedom, while adopting \( \phi(v(\vec{a})) = \sqrt{v(\vec{a})} \) shifts the focus toward its intrinsic value.

Furthermore, the compromise approach remains bounded by the two normative extremes introduced in Section~\ref{axio}. 
For any capability set~$\A$, 
the \emph{instrumental extreme} associates to~$\A$ the level set~$\A_{\max}$, 
which contains all beings in~$\mathcal{C}$ that achieve the same value as the best element of~$\A$. 
Conversely, the \emph{intrinsic extreme} associates to~$\A$ the level set~$\A_{\min}$, 
which collects all beings that attain the lowest well-being level not dominated by~$\A^\dm$. 

Together, these two level sets represent the optimistic and pessimistic interpretations of freedom. 
Formally, for any~$\A \in \mathcal{C}$, we have
\[
\Phi^{\min}_v(\A_{\min}) 
\;\leq\; 
\Phi^\phi_v(\A) 
\;\leq\; 
\Phi^{\max}_v(\A_{\max}),
\]
which guarantees that the proposed valuation~$\Phi_v$ remains \emph{bounded} between the intrinsic and instrumental extremes.

This property is illustrated in Table~\ref{tab:between_max_min}. 
For each specification of the weighting function~$\phi$, 
the integrated values obtained under the compromise evaluation lie between those generated by the intrinsic and instrumental approaches. 
Moreover, the values associated with set~$\B$, which represents the most balanced capability set with respect to the valuation function~$v$, 
are consistently closer to the bounding cases. 
In contrast, the more uneven set~$\A$ produces values that diverge more markedly from these limits.

\begin{table}[H]
\centering
\setlength{\extrarowheight}{2pt}
\begin{tabular}{|c|c|c|c|c|}
\hline
 $\phi(v(\vec{a}))=$ & Input set type & $\Phi^\phi_v(\A)$ & $\Phi^\phi_v(\B)$ & $\Phi^\phi_v(\C)$ \\ \hline
$v(\vec{a})$ 
 & Intrinsic version ($\cdot_{\min}$)   & 9 & 170.67 & 41.67 \\
 & Compromise set ($\cdot$)             & 195 & 204 & 210 \\
 & Instrumental version ($\cdot_{\max}$)& 772.33 & 243 & 576 \\ \hline
$v(\vec{a})^2$ 
 & Intrinsic version ($\cdot_{\min}$)   & 20.25 & 1024 & 156.25 \\
 & Compromise set ($\cdot$)             & 1540 & 1302 & 1475.8 \\
 & Instrumental version ($\cdot_{\max}$)& 7140.25 & 1640.25 & 5184 \\ \hline
$\sqrt{v(\vec{a})}$ 
 & Intrinsic version ($\cdot_{\min}$)   & 6.24 & 72.41 & 22.36 \\
 & Compromise set ($\cdot$)             & 74.3 & 83.95 & 83.55 \\
 & Instrumental version ($\cdot_{\max}$)& 243.73 & 97.20 & 199.48 \\ \hline
\end{tabular}
\caption{
Values of the compromise measure $\Phi^\phi_v(\cdot)$ 
for capability sets $\A$, $\B$, and $\C$, 
under different specifications of the interpretation of freedom $\phi$. 
Each row corresponds to a different input set type: 
the intrinsic version ($\cdot_{\min}$), the compromise set ($\cdot$), 
and the instrumental version ($\cdot_{\max}$)
}
\label{tab:between_max_min}
\end{table}

\subsection{Algorithms and complexity}

From a computational point of view, calculating \( \Phi^\phi_v(\A) \) presents a complex challenge that is beyond the scope of this paper. Nevertheless, we introduce Algorithm \ref{alg:integral} as a theoretical method to compute \( \Phi^\phi_v(\mathbf{A}) \) when \( P(\mathbf{A}) \) is known and finite, despite practical limitations in its application.

Considering \( \A^\dm = P(\A)^\dm \), we can express \( \Phi^\phi_v(\mathbf{A}) \) as
\[
\Phi^\phi_v(\mathbf{A}) = \Phi^\phi_v\left(\bigcup_{\vec{a} \in P(\mathbf{A})} \vec{a}\right).
\]
Applying the Principle of Inclusion-Exclusion further refines this expression to 
\[
\Phi^\phi_v\left(\bigcup_{\vec{a} \in P(\mathbf{A})} \vec{a}\right) = \sum_{Z \subseteq \mathcal{P}(P(\mathbf{A})) \setminus \{\emptyset\}} (-1)^{|Z|+1} \cdot \Phi^\phi_v\left(\bigcap_{\vec{a} \in Z} \vec{a}^\dm\right).
\]
This equation underpins the generation of all subsets of the power set of the Pareto frontier of $\A$ (line 3 of Algorithm \ref{alg:integral}).

We note that for any set \( Z \) within the power set \( \mathcal{P}(P(\mathbf{A})) \), the expression \( \bigcap_{\vec{a} \in Z} \vec{a}^\dm \) is equivalent to \( (\min_{\vec{a} \in Z} a_h \,|\, \forall h \in \{1, \ldots, h^*\})^\dm \). This equivalence forms the basis for deriving lines 4-6 of Algorithm \ref{alg:integral}. Furthermore, line 7 is developed based on the Principle of Inclusion-Exclusion. 
A crucial observation is that we transform $P(\A)$ into $P(\A)\cap\mathbb{R}^{h^*}_>$ (line 1) so that for any function \( v \in V \), the function \( \phi(v()) \) maintains strict positivity over \( \mathcal{C} \). Consequently, \( \Phi^{\phi}_v(\bigcap_{\vec{a} \in Z} \vec{a}^\dm) \) always has a non-null value.

To compute the integral indicated in line 6, we can utilize numerical methods, facilitated by the SciPy 1.0 Python package \citep{virtanen2020scipy}. While this approach is computationally intensive, it offers a viable solution for performing the required calculations in the algorithm.
It is important to note that the complexity of the integration methods in SciPy 1.0, especially the `dblquad` function, is variable and cannot be expressed using a standard fixed-complexity notation. Consequently, it is advisable to be aware that complex functions, precise requirements, and larger ranges are computationally demanding.

Furthermore, the practical feasibility of the algorithm is limited due to the exponential complexity involved in generating the power set \( \mathcal{P}(P(\mathbf{A})) \), which scales as \( O(2^{|P(\mathbf{A})|}) \). This exponential growth in complexity renders the algorithm impractical for large Pareto sets.

\begin{algorithm}[H]
\DontPrintSemicolon  
\SetAlgoLined
\LinesNumbered  
\KwIn{$P(\A)$}
\KwOut{Value of $\Phi^\phi_v(\mathbf{A})$ calculated using the compromise approach}
$P(\A)\gets P(\A)\cap\mathbb{R}^{h^*}_{>0}$\;
$\text{Output} \gets 0$\;
\ForAll{$Z \in \mathcal{P}(P(\mathbf{A}))\setminus\{\emptyset\}$}{
    \For{$h \gets 1$ \KwTo $h^*$}{
        $x_h \gets \min_{\vec{a} \in Z} a_h$\;
    }
        $\text{Output} \gets \text{Output } + (-1)^{|Z|+1} \cdot \left( \int^{x_1}_0 \cdots \right.\left. \int^{x_{h^*}}_0 \phi(v(a_1, \cdots, a_{h^*})) \, da_1 \cdots da_{h^*} \right)$}
\caption{Finding $\Phi^\phi_v(\A)$ for the compromise approach using the Inclusion-Exclusion Principle}
\label{alg:integral}
\end{algorithm}

To effectively implement this technique in scenarios where $|P(\mathbf{A})|$ is more than a few, the development of feasible approximation methods is crucial.
However, in the simple case where \( h^* = 2 \), Algorithm \ref{alg:intdim2} provides an exact method with complexity \( O(|P(\mathbf{A})\cap\mathbb{R}^{h^*}_>|) \) integration method calls. By ordering the beings on the Pareto frontier and updating the integration limits during each iteration, the algorithm partitions \( \A^\dm \) in $|P(\A)\cap\mathbb{R}^{h^*}_>|$ subsets, meaning that it creates $|P(\A)\cap\mathbb{R}^{h^*}_>|$ non empty subsets, such that each element dominated by the capability set is included in one and only one of these subsets.

\begin{algorithm}[H]
\SetAlgoLined
\DontPrintSemicolon  
\KwIn{$P(\mathbf{A})$}
\KwOut{Value of $\Phi^\phi_v(\mathbf{A})$ using the compromise approach}
$P(\A)\gets P(\A)\cap\mathbb{R}^{h^*}_{>0}$\;
$L \gets \text{Sort } P(\mathbf{A}) \text{ in lexical order prioritizing } h_1$\;
$\text{Output} \gets 0$\;
$a \gets 0$\;
\For{$\vec{b} \in L$}{
    $\text{Output} \gets \text{Output} + \int_{0}^{a_1} \int_{x}^{a_2} \phi(v(a_1, a_2)) \, da_1 \, da_2$\;
    $x \gets a_2$\;
}
\caption{Algorithm for Calculating $\Phi^\phi_v(\mathbf{A})$ via Ordered Integration when $h^*=2$}

\label{alg:intdim2}
\end{algorithm}






\section{Conclusion}

This study proposed a new approach to evaluating capability sets by integrating individual preferences over diverse options, thereby bridging the gap between negative and positive freedom within the Capability Approach.  
Our methodological contribution lies in the introduction of the \emph{compromise approach}, which preserves several desirable properties of volume-based measures \citep{xu2004ranking} while relaxing their symmetry requirement, thus allowing individual values and perspectives on freedom to be explicitly incorporated.  
Compared to the framework of \cite{gaertner2008new}, our approach additionally satisfies \emph{strong monotonicity}, ensuring that every addition of a non-dominated strictly positive being is valued.  
Future work will focus on the elicitation of the function \(\phi\), its operationalization in empirical contexts, and the development of more efficient algorithms for computing these valuations.

\section*{Acknowledgments}
The first and third author
acknowledge the support of the project CARE: Capabilities for risk Acceptability and REsilience, funded by
the Franch Research Agency (ANR) and the Regions Normandy and North France. 
We express our gratitude to David Ríos Insua, whose insights have significantly contributed to the refinement and advancement of our research.


\bibliographystyle{apalike}
\bibliography{bib}

\section*{Appendix}
\appendix
\section{Proof of Proposition~\ref{prop:ADPAcapaSets}}\label{app:proofADcapaSet}
The following lemma is the missing part in the proof of Proposition~\ref{prop:ADPAcapaSets}.
\begin{lemma}\label{lem:proofADcapaSet}
If $\A$ is bounded and closed, then $\A^{\dm}$ is bounded and closed.    
\end{lemma}
\bpr
If $\A$ is empty, the proposition is trivially true. We assume from now on that $\A$ is not empty. Clearly, $\A$ is bounded entails that $\A^{\dm}$ is bounded. In order to show that $\A^{\dm}$ is a closed set, we will show that the complement $(\A^{\dm})^c$ of $A^d$ in $\Rhnneg$ is open because it is the union of open sets. We will show that
\begin{equation}\label{eq:Admc}
(\A^{\dm})^c = \bigcup_{z \in \Rhnneg : z^{\gg} \cap \A^{\dm} = \emptyset} z^{\gg},
\end{equation}
where $z^{\gg}$ denotes the set $\{x \in \Rhnneg: x_h > z_h, \textrm{ for all } h=1, \ldots, h^*\}$.
Clearly, $(\A^{\dm})^c$ contains this union. It remains to be proven that, for all $x \in (\A^{\dm})^c \subset \Rhnneg$, there is $z$ such that $z^{\gg} \cap \A^{\dm} = \emptyset$ and $x \in z^{\gg}$.  

\ssk \noindent \textbf{Case~1: there is $y \in \A$ with $y_h >0$ for all $h = 1, \ldots, h^*$}. Let $x$ be any point of $(\A^{\dm})^c$. Consider the straight line $\{\lambda x, \lambda \in \mathbb{R}_{\geq 0} \}$. Since there is a point $y\in \A$ with all strictly positive coordinates, we have $\lambda x \leq y$, for sufficiently small values of $\lambda$. Therefore, the intersection $\{\lambda x, \lambda \in \mathbb{R}_{\geq 0} \} \cap \A^{\dm} \neq \emptyset$. Since this intersection is a bounded set,   $\sup \{\lambda x, \lambda \in \mathbb{R}_{\geq 0}, \lambda x \in \A^{\dm} \} < \infty$. Let $\ovx[\lambda] = \sup \{\lambda, \lambda \in \mathbb{R}_{\geq 0}, \lambda x \in \A^{\dm} \} $. If $\ovx[\lambda] <1$, for any $\lambda$ such that $\ovx[\lambda] < \lambda <1$, $\lambda x \not \in \A^{\dm}$ and $x \in (\lambda x)^{\gg}$. We thus have $x \in z^{\gg}$, with $z = \lambda x$.

\ssk We now prove that it is not possible that $\ovx[\lambda] = 1$ because it would imply that $x \in \A^{\dm}$, a contradiction. Let $\lambda_n, n \in \mathbb{N}$ be an increasing sequence whose limit is 1. For all $n$, consider the set $(\lambda_n x)^{\geq} \cap \A$ of points in $\A$ that dominate $\lambda_n x$. This is a non-empty closed set. Let $y_n^* \in \arg \max \{\sum_{h=1}^{h^*} y_h, y \in (\lambda_n x)^{\geq} \cap \A \}$. Such a point exists since  $\sum_{h=1}^{h^*} y_h$ is a continuous function on the compact set $(\lambda_n x)^{\geq} \cap \A$. By the Bolzano-Weierstrass theorem, there is a subsequence of the sequence $y_n^*$ that converges in the compact set $\A$. Abusing notation, we consider that $y_n^*$ denotes such a subsequence and that $\lambda_n$ are the corresponding values of $\lambda$. Let $y^*$ denote the limit of the subsequence $y_n^*$.

\msk We claim that $y^* \in \bigcap_n (\lambda_n x)^{\geq} \cap \A$. We already know that $y^* \in \A$ because $\A$ is compact. We have $y_k^* \in (\lambda_n x)^{\geq} \cap \A$, for all $k \geq n$, because $(\lambda_k x)^{\geq} \subseteq (\lambda_n x)^{\geq}$, for all $k \geq n$. Since $(\lambda_n x)^{\geq} \cap \A$ is a closed set, the sequence $y_k^*$ converges in $(\lambda_n x)^{\geq} \cap \A$, and we know that it converges to $y^*$. Hence $y^* \in (\lambda_n x)^{\geq} \cap \A$, for all $n$, and $y^* \in \bigcap_n (\lambda_n x)^{\geq} \cap \A$. This proves the claim.

\ssk It remains to be proven that $\bigcap_n (\lambda_n x)^{\geq} = x^{\geq}$. It is clear that the former contains the latter. Now, let $y' \in (\lambda_n x)^{\geq}$, for all $n$. We have that $y'_h \geq \lambda_n x_h$, for all $h = 1, \ldots, h^*$. Taking the limit for $n$ tending to $\infty$, we get $y'_h \geq x_h$, for all $h$. Hence, $y' \in x^{\geq}$.

\msk So, assuming that $\ovx[\lambda] = 1$ leads to the existence of a point $y^* \in A$ that is also in $x^{\geq}$, implying that $x \in \A^\dm$, a contradiction.

\msk \noindent \textbf{Case~2: there is no $y\in \A$ with all positive coordinates}. If $\A = \{0\}$, the lemma is trivially true. Otherwise, let $x \in (\A^\dm)^c$ and $I \subseteq \{1, \ldots, h^*\}$ be such that $x_h >0$ for $h\in I$ and $x_h =0$ for $h \notin I$. We distinguish two subcases:
\begin{itemize}
    \item There is no $y \in \A$ such that $y_h > 0$ for all $h \in I$. In this case, let $z = \lambda x$ for any $\lambda$ such that $0 < \lambda <1$. We have $x \in z^{\gg}$ and $z^{\gg} \cap \A^\dm = \emptyset$.
    \item If there is $y \in \A$ such that $y_h > 0$ for all $h \in I$, then for a sufficiently small value of $\lambda$, with $0<\lambda <1$, we have $\lambda x \leq y$ and we may apply the reasoning made in Case~1 to prove Equation~\eqref{eq:Admc}.
\end{itemize}
\epr

\section{Proof of axioms for the compromise approach}\label{app:proofsIntegralMethod}
\subsection{Strict monotonicity}
First, assume that \(\A \geq \B\). By definition of weak set dominance, for every solution in \(\B\) there is a solution in \(\A\) that weakly dominates it. This implies that the PDC of \(\B\) is a subset of the PDC of \(\A\), i.e.,
\[
\B^\dm \subseteq \A^\dm.
\]
Since the integral used to define \(\Phi^{\phi}_v\) is monotonic with respect to set inclusion, it follows that
\[
\Phi^{\phi}_v(\A) = \int_{\vec{a}\in\A^\dm} \phi\Big(v(\vec{a})\Big) \, d\vec{a}\geq \int_{\vec{b}\in\B^\dm} \phi\Big(v(\vec{b})\Big) \, d\vec{b}= \Phi^{\phi}_v(\B).
\]

\smallskip
Now assume that \(\A \gg \B\).  
By definition of strong set dominance (Definition~\ref{def:setDominance}), there exist a being \(\vec{a}_0 \in \A\) and \(\varepsilon > 0\) such that the hyper-rectangle
\([\vec{a}_0 - \varepsilon \cdot \vec{1}, \vec{a}_0]\) is contained in \(\Rhpos\) and does not intersect \(\B^\dm\), i.e.,
\[
[\vec{a}_0 - \varepsilon \cdot \vec{1}, \vec{a}_0] \cap \B^\dm = \emptyset.
\]
Hence, this $\varepsilon$-cube is a measurable subset of
\(\A^\dm \setminus \B^\dm\) with strictly positive Lebesgue measure.

\smallskip
We can thus decompose the integral as follows:
\[
\A^\dm = \B^\dm \cup (\A^\dm \setminus \B^\dm),
\]
and by the additivity of the integral we obtain
\[
\Phi^{\phi}_v(\A)
= \int_{\vec{b}\in\B^\dm} \phi\!\big(v(\vec{b})\big) \, d\vec{b}
  + \int_{\vec{a}\in\A^\dm \setminus \B^\dm} \phi\!\big(v(\vec{a})\big) \, d\vec{a}.
\]

Since \(\phi(v(\vec{a}))\) is strictly positive on its domain  
and the set \(\A^\dm \setminus \B^\dm\) has positive measure  
(because it contains the nonempty cube \([\vec{a}_0 - \varepsilon \cdot \vec{1}, \vec{a}_0]\)),  
the second integral is strictly positive:
\[
\int_{\vec{a}\in\A^\dm \setminus \B^\dm} \phi\!\big(v(\vec{a})\big) \, d\vec{a} > 0.
\]
Therefore,
\[
\Phi^{\phi}_v(\A) > \Phi^{\phi}_v(\B).
\]
\hfill\(\blacksquare\)

\subsection{Invariance of Scaling Effects} 
When we scale the set \(\A\) by \(\vec{\alpha}\), each vector \(\vec{a}\) in \(\A\) is mapped to \(\vec{a}' = \vec{\alpha} \cdot \vec{a}\). A standard property of Lebesgue measure in \(\mathbb{R}^{h^*}_>\) is that scaling every coordinate by a positive factor multiplies the measure of any measurable set by a constant. In particular, there exists a constant
\[
c = \alpha_1\alpha_2 \cdots \alpha_{h^*} > 0,
\]
such that for any integrable function \(f\) and measurable set \(\A^\dm\) we have
\[
\int_{\vec{a^\prime}\in\vec{\alpha}\cdot\A^\dm} f(\vec{a}')\, d\vec{a}' = c \int_{\vec{a}\in\A^\dm} f(\vec{a})\, d\vec{a}.
\]

In our setting, the definition of \(v'\) ensures that
\[
\phi\Big(v'\bigl(\vec{\alpha}\cdot\vec{a}\bigr)\Big) = \phi\Big(v(\vec{a})\Big),
\]
so that the evaluation function for the scaled set becomes
\[
\Phi^{\phi}_{v'}(\A') = c\, \Phi^{\phi}_v(\A),
\]
and similarly,
\[
\Phi^{\phi}_{v'}(\B') = c\, \Phi^{\phi}_v(\B).
\]
Since \(c\) is a positive constant and by hypothesis \(\Phi^{\phi}_v(\A) \geq \Phi^{\phi}_v(\B)\), it follows immediately that
\[
\Phi^{\phi}_{v'}(\A') \geq \Phi^{\phi}_{v'}(\B').
\]
$\hfill\blacksquare$

\subsection{Continuity}
By monotonicity of the integral, we have
\[
\Phi^{\phi}_v(\A^\dm\cap\C^\dm) \le \Phi^{\phi}_v(\C) < \Phi^{\phi}_v(\A) \le \Phi^{\phi}_v(\A^\dm\cup\C^\dm).
\]
For \(t\) in the interval
\[
\Bigl[0,\; \max_{\vec{x}\in \A^\dm\cup\C^\dm}v(\vec{x})\Bigr],
\]
define the function
\[
G(t)= \int_{\vec{x}\in \A^\dm\cap\C^\dm}\phi\bigl(v(\vec{x})\bigr)d\vec{x} \;+\; \int_{\substack{\vec{x}\in (\A^\dm\cup\C^\dm)\setminus(\A^\dm\cap\C^\dm) \\ v(\vec{x})\le t}}\phi\bigl(v(\vec{x})\bigr)d\vec{x}.
\]
Because \(\phi\) is strictly positive, $\phi$ and $v$ are continuous, and the integral is monotone \wrt the integration domain, \(G(t)\) is continuous and non decreasing with respect to \(t\). In particular, when \(t\) is minimal (i.e., \(t=0\)) we have
\[
G(0)= \int_{\vec{x}\in \A^\dm\cap\C^\dm}\phi\bigl(v(\vec{x})\bigr)d\vec{x} = \Phi^{\phi}_v(\A^\dm\cap\C^\dm),
\]
and when \(t\) is maximal (i.e., \(t=\max_{\vec{x}\in \A^\dm\cup\C^\dm} v(\vec{x})\))
\[
G\Bigl(\max_{\vec{x}\in \A^\dm\cup\C^\dm} v(\vec{x})\Bigr)= \int_{\vec{x}\in \A^\dm\cup\C^\dm}\phi\bigl(v(\vec{x})\bigr)d\vec{x} = \Phi^{\phi}_v(\A\cup\C).
\]
Since
\[
\Phi^{\phi}_v(\A\cap\C) \le \Phi^{\phi}_v(\C) < \lambda < \Phi^{\phi}_v(\A) \le \Phi^{\phi}_v(\A\cup\C),
\]
the Intermediate Value Theorem guarantees that there exists a \(t^*\) in
\[
\Bigl[0,\; \max_{\vec{x}\in \A^\dm\cup\C^\dm}v(\vec{x})\Bigr]
\]
such that
\[
G(t^*) = \lambda.
\]
Now, define the set
\[
\B=\Bigl\{ \vec{x}\in \A^\dm\cup\C^\dm :\; \vec{x}\in \A^\dm\cap\C^\dm \;\text{ or }\; v(\vec{x})\le t^*\Bigr\}.
\]
Clearly, \(\A^\dm\cap\C^\dm\subseteq \B\subseteq \A^\dm\cup\C^\dm\). By construction, the evaluation of \(\B\) is given by
\[
\Phi^{\phi}_v(\B) = G(t^*) = \lambda.
\]
This completes the proof.$\hfill\blacksquare$

\subsection{Bounded Freedom Principle}
Proof by monotonicity since we have $\A_{\min}^\dm\subseteq\A^\dm\subseteq\A^\dm_{\max}$
$\hfill\blacksquare$
\section{Proof of Proposition~\ref{prop:eqVol}}\label{app:eqVol}
If \(\phi\) is constant with value \(c\), then for all \(\vec{a} \in \A^\dm\), \(\phi(v(\vec{a})) = c\). Substituting this into the definition of \(\Phi^{\phi}_v\), we have
\[
\Phi^{\phi}_v(\A) = \int_{\vec{a} \in \A^\dm} \phi(v(\vec{a})) \, d\vec{a} = \int_{\vec{a} \in \A^\dm} c \, d\vec{a} = c \int_{\vec{a} \in \A^\dm} 1 \, d\vec{a}.
\]
By definition, \(\int_{\vec{a} \in \A^\dm} 1 \, d\vec{a} = \mathrm{vol}(\A^\dm)\). Therefore,
\[
\Phi^{\phi}_v(\A) = c \cdot \mathrm{vol}(\A^\dm).
\]
$\hfill\blacksquare$

\end{document}